\documentclass[11pt]{article}

\usepackage{amsmath,amssymb,amsfonts}
\usepackage{amsthm}
\usepackage{bm}
\usepackage{mathtools}
\usepackage{geometry}
\usepackage{enumitem}
\usepackage[numbers,sort&compress]{natbib}
\usepackage{hyperref}

\usepackage{algorithm}
\usepackage{algpseudocode}

\usepackage{tabularx}
\usepackage{booktabs}

\usepackage{tikz}
\usetikzlibrary{positioning,arrows.meta}

\newcommand\ci{\perp\!\!\!\perp}

\newcommand{\E}{\mathbb{E}}

\newtheorem{assumptions}{Assumptions}

\title{Causal Outcome Learning with Noisy Proxies}
\author{Eric V. Strobl\\
University of Pittsburgh}
\date{}

\begin{document}

\maketitle

\begin{abstract}
\noindent\textbf{Objective:}
Modern observational biomedical datasets contain increasingly extensive baseline and outcome measurements, yet detecting a clinically meaningful treatment effect remains difficult because investigators must identify an outcome that captures treatment response while adjusting for confounding distributed across numerous noisy baseline variables. We introduce SCOUT (Score-based Causal Outcome Learning Using Treatment signals), a framework that learns a clinically interpretable, treatment-responsive outcome while adjusting for confounding throughout the learning process.

\noindent\textbf{Methods:}
SCOUT constructs low-dimensional adjustment representations of high-dimensional baseline measurements to estimate the average treatment effect (ATE) and conditional average treatment effect (CATE) for the learned outcome. Importantly, the final adjustment space is at most two-dimensional, allowing treatment-effect heterogeneity to be estimated in this space rather than in the full high-dimensional covariate space. SCOUT also introduces a fourth-order statistic for testing treatment-effect heterogeneity that emphasizes unusually large deviations from the ATE.

\noindent\textbf{Results:}
In synthetic and semi-synthetic experiments, the algorithm achieved competitive ATE estimation accuracy and substantially lower CATE estimation error. In an observational study of patients with bipolar disorder, SCOUT identified a clinically recognizable lithium-associated adverse-effect profile, detecting both a significant ATE (Holm-adjusted $p$ = 0.0026) and significant treatment-effect heterogeneity (Holm-adjusted $p$ = 0.033).

\noindent\textbf{Conclusion:}
These results demonstrate how jointly learning outcomes and controlling for confounding can facilitate the discovery of treatment effects in high-dimensional observational data.
\end{abstract}


\section{Introduction}

Estimating causal effects from high-dimensional biomedical observational datasets is challenging, even with large sample sizes. A central difficulty is identifying an appropriate set of baseline variables for confounding adjustment. Investigators often lack sufficient knowledge to isolate this variable set, particularly when confounding information is distributed across many noisy variables rather than captured by a small set of directly measured confounders \cite{Hong19}. Many have responded by adjusting for increasingly large numbers of covariates, a tendency reinforced by the growing availability of additional variables in modern datasets \cite{zhang2022lsps}. However, indiscriminate adjustment can exacerbate violations of the positivity assumption and explain most of the variation in treatment assignment, leaving little residual treatment variation for estimating causal effects \cite{d2021overlap}.

The large number of variables collected in modern observational studies creates another challenge: selecting an appropriate outcome. When investigators fail to detect treatment effects, they may question whether the chosen outcome adequately captures treatment response. Datasets often contain measurements of numerous symptoms, laboratory values, functional outcomes, and disease features, yet conventional causal analyses typically require investigators to prespecify a single endpoint or combine multiple measurements into a fixed composite score. This can obscure treatment effects when only a subset of outcome components responds or when response follows a particular multivariate pattern among the outcomes \cite{strobl2026taskaligned,strobl2025score}. Moreover, testing multiple candidate outcomes separately introduces multiple-testing penalties that further reduce statistical power. Investigators thus face a fundamental dilemma: clinically meaningful treatment effects may be difficult to detect using any single prespecified outcome, yet identifying these effects also requires adjusting for potentially numerous confounders. As the number of adjustment variables increases, the resulting loss of statistical power can even offset the benefits of larger sample sizes in observational datasets \citep{myers2011effects}.

The above challenges become even more pronounced when investigators seek to identify treatment effect heterogeneity. Beyond detecting potentially small causal effects while adjusting for numerous confounders, investigators must also identify systematic differences in treatment response across individuals. Detecting such fine-grained heterogeneity typically requires large sample sizes to achieve adequate statistical power, often exceeding those available in practice \citep{brookes2004subgroup}. Indeed, a reanalysis of 32 large randomized trials found statistically significant average treatment effects (ATEs) in 18 trials, yet only one of these 18 (5.6\%) also demonstrated a statistically significant interaction between treatment and baseline risk \citep{kent2020predictive}. Thus, even studies that detect significant ATEs frequently fail to establish statistically significant heterogeneity in conditional average treatment effects (CATEs).

We need methods that can accommodate high-dimensional confounding adjustment and multiple candidate outcomes, even when treatment effects on individual outcomes are small. We introduce such an algorithm called SCOUT (\textbf{S}core-based \textbf{C}ausal \textbf{O}utcome learning \textbf{U}sing \textbf{T}reatment signals), which learns a treatment-responsive outcome composite while reducing high-dimensional baseline information to a low-dimensional representation for confounding adjustment. SCOUT further estimates CATEs within this low-dimensional space, improving estimation accuracy and facilitating the detection of treatment-effect heterogeneity.

We specifically make the following contributions:

\begin{enumerate}[leftmargin=1.25em,label=(\alph*),labelsep=0.4em,itemsep=0.25em]

\item We develop a method for learning treatment-responsive composite outcomes from multiple longitudinal measurements rather than relying on prespecified endpoints.

\item We construct low-dimensional adjustment scores from high-dimensional, noisy baseline variables to control for confounding throughout the outcome-learning process.

\item We further reduce the adjustment scores to just two dimensions after outcome learning to accurately estimate treatment effects.

\item We develop conditional randomization tests (CRTs) for average and heterogeneous treatment effects that account for adaptive outcome learning. For heterogeneity testing, we use a fourth-order statistic that emphasizes patients with unusually large treatment responses, reflecting the clinical importance of identifying individuals who respond particularly well or poorly.

\item We evaluate SCOUT through synthetic and semi-synthetic experiments, demonstrating competitive ATE accuracy, substantially lower CATE estimation error, and high computational efficiency relative to existing approaches. We further illustrate its clinical utility in an observational study of lithium-associated adverse effects in patients with bipolar disorder.

\end{enumerate}
Together, these contributions establish SCOUT as a unified framework for learning clinically meaningful outcomes, estimating average and heterogeneous causal treatment effects, and testing their statistical significance in high-dimensional observational data.

\section{Related Work}

Existing methods address high-dimensional confounding, noisy proxy variables, and multivariate outcome construction largely as separate problems. Classical propensity-score and doubly robust estimators focus on adjustment for measured confounding while treating the outcome as prespecified \citep{rosenbaum1983central,bang2005doubly}. Extensions to high-dimensional settings use regularization, variable screening, or automated covariate selection to make adjustment feasible when the number of candidate variables is large \citep{schneeweiss2009hdps,gao2022highdim}. These approaches improve confounding control but do not address uncertainty about how the clinical outcome itself should be defined. Moreover, measurement error in adjustment variables can leave residual bias when the underlying confounders are not adequately recovered \citep{Hong19}.

A complementary literature learns outcomes from multiple endpoints. Zhou et al.\ use cross-validated targeted maximum likelihood estimation to learn composite-outcome weights that increase power in randomized clinical trials \citep{zhou2026cvtmle}. Inverse regression similarly provides data-driven weights for multiple outcomes and can be applied in both randomized and observational studies, but it optimizes standardized treatment separation rather than a clinical direction such as coordinated symptom improvement \citep{zhang2025inverse}. Luckett et al.\ learn composite-outcome weights by identifying the utility function whose implied optimal treatment decisions best explain treatment assignments in observational data \citep{luckett2021estimation}. SCORE learns sparse longitudinal outcome combinations that maximize predictability, but does not target causal treatment effects \citep{strobl2025score}. Supervised Varimax instead learns outcomes that maximally differentiate treatments, primarily addressing comparative treatment effects rather than confounding from nonrandom treatment assignment \citep{stroblkim2026sv}. Thus, these methods demonstrate the value of learning a clinically interpretable endpoint in different contexts, but do not simultaneously solve the proxy-adjustment problem arising in observational studies.

Conversely, methods for adjusting for latent confounding using observed proxies focus on recovering sufficient adjustment information from imperfect measurements of unobserved confounders, typically treating the outcome as prespecified. Proximal causal inference permits causal identification when measured variables are imperfect proxies for latent confounders, but requires variables satisfying distinct treatment- and outcome-proxy roles, together with bridge-function and completeness conditions \citep{tchetgen2024proximal}. The large-scale propensity score approach exploits large collections of observed variables to account for latent confounding factors, but its theoretical guarantee of eliminating confounding bias relies on a strong pinpointability assumption that the latent confounders are deterministic functions of the observed covariates \citep{zhang2022lsps}. More recently, the COMPACT algorithm relaxes this requirement by allowing approximate recovery of confounding information \citep{strobl2026compact}. It further exploits the underlying causal structure to learn a low-dimensional adjustment representation that maximizes recovery of the relevant confounding information from high-dimensional, noisy, and overlapping baseline proxies. Its current formulation, however, only considers a prespecified scalar outcome.

The distinguishing contribution of SCOUT is consequently not any one of these components in isolation, but their integration. Unlike methods that address adjustment or outcome learning separately, SCOUT constructs a fixed, low-dimensional representation of high-dimensional, noisy baseline variables that enables confounding adjustment throughout the outcome-learning process. This allows SCOUT to efficiently learn treatment-responsive composite outcomes without repeatedly estimating the adjustment representation.

\section{Methods}

\subsection{Notation}

We consider $n$ individuals, each with a binary treatment
$T\in\{0,1\}$, an unobserved $q$-dimensional pretreatment
state $\bm C\in\mathbb R^q$ representing latent confounders,
and a $p$-dimensional vector of observed pretreatment variables
$\bm X\in\mathbb R^p$ that serve as potentially noisy and
redundant proxies for $\bm C$ (Figure \ref{fig:causal_dag}).
For each individual, an $r$-dimensional clinical outcome vector
may be observed at each of $m$ post-treatment visits. Let
$\mathcal I_k\subseteq\{1,\ldots,n\}$ denote the set of individuals
observed at visit $k$, and let $n_k=|\mathcal I_k|$.
Let $\mathbf X\in\mathbb R^{n\times p}$ denote the full baseline
matrix and $\mathbf T=(T_1,\ldots,T_n)^\top$ the full treatment vector.
Write $\mathbf X_k=\mathbf X[\mathcal I_k,:]$ and
$\mathbf T_k=\mathbf T[\mathcal I_k]$ for their visit-specific subsets.
Order the indices in $\mathcal I_k$ as $i_1,\ldots,i_{n_k}$,
consistently across the visit-specific matrices. Let
\[
\mathbf{Y}_k=
\begin{pmatrix}
\bm Y_{i_1k}^{\top}\\
\vdots\\
\bm Y_{i_{n_k}k}^{\top}
\end{pmatrix}
\in\mathbb R^{n_k\times r},
\qquad
k=1,\ldots,m,
\]
denote the observed outcome matrix at visit $k$, where $\bm Y_{ik}$
contains the $r$ clinical outcome components for individual $i$.
We will describe the causal assumptions and underlying model in detail
in the following subsections.

Before analysis, we orient outcome components according to the treatment-response pattern of interest. For example, symptom outcomes may be oriented so that larger values indicate greater improvement, while adverse-effect outcomes may be oriented so that larger values indicate greater adverse-effect severity, allowing SCOUT to identify treatment-response profiles involving both therapeutic benefits and adverse effects. Components are multiplied by $-1$ when necessary to achieve the prespecified orientation. We code $T=1$ as the treatment of interest and $T=0$ as the comparator. Consequently, a positive treatment effect indicates greater expression of the learned outcome profile in the prespecified directions.

SCOUT will learn an outcome weight vector $\bm\alpha \geq 0$. For any candidate weight vector, define the corresponding phenotype at visit $k$ as $\mathbf Y_k^{*}(\bm\alpha)=\mathbf{Y}_k\bm\alpha$. Let $\mathbf Y_k^{*}(t;\bm\alpha)$ denote its potential value under treatment state $t\in\{0,1\}$. After learning $\widehat{\bm\alpha}$, the learned phenotype is therefore $\mathbf Y_k^{*}=\bm{Y}_k\widehat{\bm\alpha}$.

\subsection{Causal assumptions}

\begin{figure}[!b]
\centering
\begin{tikzpicture}[
    >=stealth,
    every node/.style={font=\small}
]

\node (C) at (0,1.5) {$\bm C$};
\node (X) at (-1.35,0) {$\bm X$};
\node (T) at (1.05,0) {$T$};
\node (Y1) at (3.0,0) {$\bm Y_1$};
\node (Y2) at (4.875,0) {$\bm Y_2$};
\node (dots) at (6.225,0) {$\cdots$};
\node (Ym) at (7.575,0) {$\bm Y_m$};

\draw[->] (C) -- (X);
\draw[->] (C) -- (T);
\draw[->] (C) -- (Y1);
\draw[->] (C) -- (Y2);
\draw[->] (C) -- (Ym);

\draw[->,dashed] (T) -- (Y1);
\draw[->,dashed] (T) to[bend right=15] (Y2);
\draw[->,dashed] (T) to[bend right=17] (Ym);
\draw[->,dashed] (Y1) -- (Y2);

\end{tikzpicture}
\caption{Schematic of the assumed causal relationships.
The latent pretreatment state $\bm C$ influences treatment
assignment $T$, the observed baseline proxies $\bm X$,
and the longitudinal outcomes $\bm Y_1,\ldots,\bm Y_m$.
Dashed arrows indicate possible causal effects, which
are not assumed to be nonzero. }
\label{fig:causal_dag}
\end{figure}
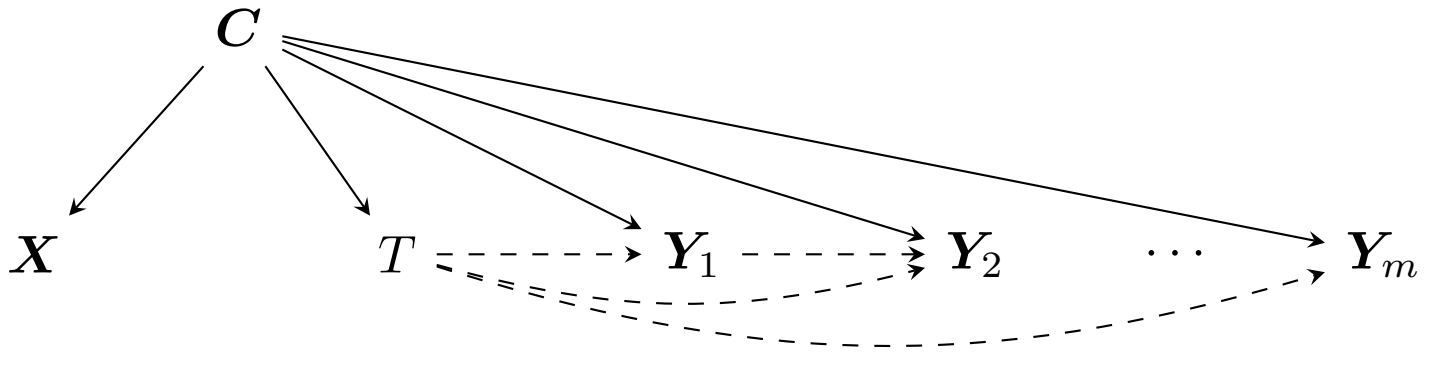

We adopt the following standard causal inference assumptions, formulated in terms of the latent confounders $\bm C$ rather than their observed proxies $\bm X$, consistent with the causal structure illustrated in Figure \ref{fig:causal_dag}.

\begin{assumptions}
The following conditions hold:

\begin{enumerate}[leftmargin=*]

\item \textbf{Consistency and no interference.}
For each individual $i$, treatment state $t\in\{0,1\}$,
visit $k$, and fixed outcome weight vector $\bm\alpha$,
\[
T=t
\quad\Longrightarrow\quad
Y_{ik}^{*}(\bm\alpha)
=
Y_{ik}^{*}(t;\bm\alpha).
\]
In addition, individual $i$'s potential outcomes do not depend on the treatment assignments of other individuals.

\item \textbf{Conditional exchangeability.}
There exists a pretreatment latent baseline state $\bm C$ sufficient for confounding adjustment such that, for every visit $k$ and fixed $\bm\alpha$,
\[
\left\{
Y_{ik}^{*}(0;\bm\alpha),
Y_{ik}^{*}(1;\bm\alpha)
\right\}
\ci T\mid\bm C.
\]

\item \textbf{Positivity.}
For all $\bm c$ in the support of $\bm C$,
\(
0<
\mathbb P(T=1\mid\bm C=\bm c)
<1.
\)

\end{enumerate}
\end{assumptions}

\subsection{Noisy-proxy and treatment-effect model}

In prior work, we showed that, in the case of a scalar treatment and scalar outcome, confounding adjustment can be achieved using at-most two learned scores rather than the unobserved confounders $\bm C$, under conditions concerning how the latent confounders affect treatment and outcomes and how well the observed proxies capture this information \citep{strobl2026compact}. Let $\bm\gamma_k(\bm\alpha)\in\mathbb R^{p\times d}$, with $d\leq2$, denote the coefficient matrix for the learned scores. The score matrix for visit $k$, evaluated for all $n$ individuals, is
\(
\mathbf S_k(\bm\alpha)=\mathbf X\bm\gamma_k(\bm\alpha),
\)
whose row for individual $i$ is $\bm S_{k,i}(\bm\alpha)^\top$. The scores are learned from individuals observed at visit $k$, but can be applied to the baseline covariates of all individuals. The key assumptions are that treatment assignment depends on a single linear combination of the unobserved confounders, potential outcomes depend linearly on the confounders, and the observed proxies collectively contain sufficient information to recover the score variables. We refer readers to the original paper for the precise conditions and theoretical guarantees. SCOUT builds on this framework to learn the outcome itself while accounting for confounding.

For a given outcome weight vector $\bm\alpha$, we define the score-indexed CATE as
\[
\tau_k(\bm s;\bm\alpha)
=
\E\!\left[
Y_{ik}^{*}(1;\bm\alpha)
-
Y_{ik}^{*}(0;\bm\alpha)
\mid \bm S_{k,i}(\bm\alpha)=\bm s
\right].
\]
This estimand describes treatment-effect heterogeneity within
the learned adjustment space rather than over high dimensional $\bm{X}$. The corresponding
ATE is
\(
\bar\tau_k(\bm\alpha)
=
\E_{\bm S_{k,i}(\bm\alpha)}
\left[
\tau_k\!\left(\bm S_{k,i}(\bm\alpha);\bm\alpha\right)
\right].
\)

We assume the linear potential outcomes model
\begin{equation}\label{eq:causal_model}
Y_{ik}^{*}(\bm\alpha)
=
\beta_{0k}(\bm\alpha)
+
\beta_{T,k}(\bm\alpha)T
+
\bm\beta^{\top}_{S,k}(\bm\alpha)\bm S_{k,i}(\bm\alpha)
+
T\bm\beta^{\top}_{TS,k}(\bm\alpha)\bm S_{k,i}(\bm\alpha)
+
\varepsilon_{ik}(\bm\alpha),
\end{equation}
where $\E\!\left[
\varepsilon_{ik}(\bm\alpha)
\mid T,\bm S_{k,i}(\bm\alpha)
\right]=0$. Thus, the score-indexed CATE is more specifically
\[
\tau_k(\bm s;\bm\alpha)
=
\beta_{T,k}(\bm\alpha)
+
\bm\beta^{\top}_{TS,k}(\bm\alpha)\bm s,
\]
and the corresponding ATE is
\[
\bar\tau_k(\bm\alpha)
=
\beta_{T,k}(\bm\alpha)
+
\bm\beta^{\top}_{TS,k}(\bm\alpha)
\E\!\left[\bm S_{k,i}(\bm\alpha)\right].
\]

\subsection{SCOUT Algorithm}

\subsubsection{Initial adjustment representation}

Prior work showed that, for a scalar treatment and scalar outcome,
confounding adjustment can be achieved using at most two adjustment
scores under appropriate conditions \citep{strobl2026compact}.
However, SCOUT considers multiple outcomes and learns a composite
from them, so the two scores constructed for one candidate composite
may not be sufficient for another. We therefore construct a common
set of initial adjustment scores that can simultaneously represent
the two adjustment scores needed for every possible outcome composite.

At each visit $k$, define the centering matrix
\(
\mathbf{H}_{n_k}=\mathbf{I}_{n_k}-\frac{1}{n_k}\bm 1\bm 1^\top.
\)
Write $\mathbf{X}_{c,k}=\mathbf{H}_{n_k}\mathbf{X}_k$ and
$\mathbf{T}_{c,k}=\mathbf{H}_{n_k}\mathbf{T}_k$. The operator
\(
\mathbf{M}_{T,k}
=
\mathbf{H}_{n_k}-
\mathbf{T}_{c,k}
(\mathbf{T}_{c,k}^\top\mathbf{T}_{c,k})^{-1}
\mathbf{T}_{c,k}^\top
\)
removes variation explained by the intercept and treatment.
Define the treatment direction
\(
\bm v_{T,k}=n_k^{-1}\mathbf{X}_{c,k}^\top\mathbf{T}_{c,k}
\)
and the regularized matrix
\[
\mathbf{G}_{k,\eta}
=
\frac{1}{2n_k}
\left(
\mathbf{X}_{c,k}^\top\mathbf{X}_{c,k}+
\mathbf{X}_{c,k}^\top\mathbf{M}_{T,k}\mathbf{X}_{c,k}
\right)
+\eta\mathbf{I}_p,
\]
where $\eta>0$ is a ridge parameter optimized by cross-validation.

At visit $k$, define the treatment-residualized
outcome-direction matrix
\[
\mathbf{V}_{R,k}
=
n_k^{-1}\mathbf{X}_{c,k}^\top\mathbf{M}_{T,k}\mathbf{Y}_k
=
\begin{bmatrix}
\bm v_{R,k1} & \cdots & \bm v_{R,kr}
\end{bmatrix}.
\]
Following \citep{strobl2026compact}, for a given
nonnegative outcome weight vector $\bm\alpha$, the two
adjustment scores for the composite $\mathbf{Y}_k\bm\alpha$
can be represented by
\[
\mathbf{S}_{k,\eta}(\bm\alpha)
=
\mathbf{X}
\left[
\mathbf{G}_{k,\eta}^{-1}\bm v_{T,k},\;
\mathbf{G}_{k,\eta}^{-1}\mathbf{V}_{R,k}\bm\alpha
\right].
\]
Importantly, both columns of
$\mathbf{S}_{k,\eta}(\bm\alpha)$ are linear combinations
of the columns of
\[
\mathbf{S}_{k}^{(0)}
=
\mathbf{X}
\left[
\mathbf{G}_{k,\eta}^{-1}\bm v_{T,k},\;
\mathbf{G}_{k,\eta}^{-1}\mathbf{V}_{R,k}
\right].
\]
The first column corresponds to treatment, while the
remaining $r$ columns correspond to the individual outcomes.
Because any candidate composite is a nonnegative linear
combination of these outcomes, its adjustment scores can
be recovered from $\mathbf{S}_{k}^{(0)}$.
Thus, SCOUT requires at most $r+1$ adjustment
scores at each visit to accommodate all candidate outcome composites.
Crucially, these scores can be computed once and held
fixed throughout outcome learning, because the two
adjustment scores for any candidate composite can be
recovered as linear combinations of them. 

\subsubsection{Learning the composite outcome}

Using the initial adjustment scores constructed in the preceding
subsection, SCOUT residualizes the outcome matrix $\mathbf{Y}_k$
and treatment vector $\mathbf{T}_k$ with respect to
the rows of $\mathbf{S}_k^{(0)}$ for individuals in $\mathcal I_k$, yielding $\widetilde{\mathbf{Y}}_k$
and $\widetilde{\mathbf{T}}_k$, respectively. The algorithm then learns a single nonnegative outcome weight vector shared
across all follow-up visits by solving the nonnegative
least-squares (NNLS) problem
\begin{equation} \label{eq:NNLS}
\widehat{\bm\alpha}
=
\arg\min_{\bm\alpha\geq\bm 0}
\sum_{k=1}^{m}
\frac{1}{n_k}
\left\|
\widetilde{\mathbf{T}}_k
-
\widetilde{\mathbf{Y}}_k\bm\alpha
\right\|_2^2,
\end{equation}
Normalizing by $n_k$
gives each follow-up visit equal weight in the objective,
preventing visits with more observations from dominating
the optimization solely because of their sample size. For interpretability, we normalize the learned outcome weights to sum to one after optimization, provided that at least one weight is nonzero.

The nonnegativity constraint $\bm\alpha\geq\bm 0$ is important for clinical interpretability because all outcome components are oriented in prespecified clinical directions. Without this constraint, regression may assign negative weights even to outcomes positively associated with treatment after adjustment, because the coefficients depend on the covariance structure among the residualized outcomes and may exhibit suppression effects.

\subsubsection{Adjustment after outcome learning}

The initial $r+1$ adjustment scores at each visit are constructed to accommodate all candidate outcome composites during NNLS optimization. Once $\widehat{\bm\alpha}$ has been determined, the problem reduces to the scalar-treatment, scalar-outcome setting, where at most two adjustment scores are sufficient under the previously described conditions.

We therefore fix $\widehat{\bm\alpha}$ and construct the learned scalar outcome from the original, unresidualized clinical measurements:
\(
\mathbf Y_k^{*}=\mathbf{Y}_k\widehat{\bm\alpha}.
\)
SCOUT then relearns two-dimensional adjustment scores separately at each visit using the original data $(\mathbf{X}_k,\mathbf{T}_k,\mathbf Y_k^{*})$. Define
\(
\bm v_{R,k}^{*}
=
n_k^{-1}\mathbf{X}_k^{\top}\mathbf{M}_{T,k}\mathbf Y_k^{*}.
\)
Define the visit-specific score-direction matrix
\[
\mathbf D_k
=
\left[
\mathbf{G}_{k,\eta}^{-1}\bm v_{T,k},\;
\mathbf{G}_{k,\eta}^{-1}\bm v_{R,k}^{*}
\right].
\]
The resulting two-dimensional adjustment scores are
\(
\mathbf S_k=\mathbf X\mathbf D_k.
\)
These refined scores are then used for the final estimation of average and heterogeneous treatment effects.

\subsubsection{ATE and CATE estimation}

At each visit $k$, SCOUT finally fits the treatment-effect model in \eqref{eq:causal_model} by ordinary least squares using the rows of the targeted scalar scores $\mathbf{S}_k$ for individuals in $\mathcal I_k$. For each $i\in\mathcal I_k$, the resulting subject-specific CATE estimate is then
\(
\widehat\tau_{ik}
=
\widehat\beta_{T,k}
+
\widehat{\bm\beta}^{\top}_{TS,k} \bm S_{k,i}.
\) Let $\overline{\bm S}_k$ denote the mean of the final adjustment scores among individuals in $\mathcal I_k$. Averaging these subject-specific effects gives the empirical ATE,
\[
\widehat{\bar\tau}_k
=
\frac{1}{n_k}
\sum_{i\in\mathcal I_k}
\widehat\tau_{ik}
=
\widehat\beta_{T,k}
+
\widehat{\bm\beta}^{\top}_{TS,k}\overline{\bm S}_k.
\]
We provide pseudocode of the SCOUT algorithm in Supplementary Material \ref{supp:sec:pseudocode} and complexity analysis in Supplementary Material \ref{app:complexity}. For a fixed ridge parameter
$\eta$, SCOUT scales linearly with the number of baseline
individuals $n$ and the total number of observed
individual-visit observations $N=\sum_{k=1}^{m}n_k$,
cubically with the number of baseline variables
$p$, quadratically with the number of outcome
components $r$, and linearly with
the number of follow-up visits $m$. In practice, the algorithm completed in at most 1.2 seconds on average in our experiments.

\subsection{Hypothesis testing}

We use a CRT that repeats the complete SCOUT procedure under randomized treatment assignments to account for adaptive outcome and adjustment-score construction \citep{candes2018panning}. Because each $\mathbf S_k$ contains scores for all $n$ individuals, we estimate a single propensity model $\widehat{\mathbb P}(T_i=1\mid \bm S_{1,i},\ldots,\bm S_{m,i})$ by logistic regression using the concatenated $2m$ scores. We then generate $B$ randomized treatment vectors from the estimated propensity model, using the same randomized treatment assignment for each individual across all visits. We refit SCOUT in each randomization.

\subsubsection{ATE test}
Because SCOUT's NNLS objective favors outcome composites positively associated with treatment, we test the conditional-independence null
$H_0^{\mathrm{ATE}}:
T \ci (\bm{Y}_1,\ldots,\bm{Y}_m)
\mid (\bm{S}_1,\ldots,\bm{S}_m)$
against conditional dependence using the estimated ATE
$\widehat{\bar\tau}_k$ as a one-sided test statistic.
Let $\widehat{\bar\tau}_k^{(b)}$ denote the corresponding estimate
from randomization $b$. The upper-tail $p$-value is
\[
p_k^{\mathrm{ATE}}
=
\frac{
1+\sum_{b=1}^{B}
\mathbf{1}\{
\widehat{\bar\tau}_k^{(b)}
\geq
\widehat{\bar\tau}_k
\}
}{
B+1
}.
\]
We also perform a pooled test using the mean of the visit-specific ATE estimates as the test statistic.

\subsubsection{CATE heterogeneity test}
For treatment-effect heterogeneity, we test the null hypothesis that treatment is conditionally independent of the outcomes after removing constant outcome-specific treatment effects,
\[
H_0^{\mathrm{CATE}}:
\exists\,\bm\delta_1,\ldots,\bm\delta_m
\text{ such that }
T \ci
(\bm Y_1-T\bm\delta_1,\ldots,\bm Y_m-T\bm\delta_m)
\mid
(\bm S_1,\ldots,\bm S_m).
\]
This null allows for a nonzero ATE. We therefore estimate a vector of constant outcome-specific treatment effects $\widehat{\bm\delta}_k$ by regressing the outcome components on treatment and their corresponding adjustment scores. Let $\mathbf T_k^{(b)}$ denote the entries of the randomized treatment vector for individuals in $\mathcal I_k$. For each randomization $b$, we construct
\(
\mathbf{Y}_k^{(b)}
=
\mathbf{Y}_k
+
\left(
\mathbf{T}_k^{(b)}-\mathbf{T}_k
\right)
\widehat{\bm\delta}_k^\top.
\)
This transformation preserves the estimated constant treatment effects while changing the treatment assignments. SCOUT is then refitted using the randomized treatment and transformed outcomes at every visit.

Note that clinically important heterogeneity is typically concentrated among a relatively small subgroup of patients who experience unusually strong benefits or adverse effects \citep{kent2016risk}. We therefore use a fourth-order statistic that places greater weight on large deviations:
\[
Q_k
=
\sqrt{
\frac{1}{n_k}
\sum_{i\in\mathcal I_k}
\left(
\widehat\tau_{ik}
-
\widehat{\bar\tau}_k
\right)^4
}.
\]
This statistic emphasizes patients whose treatment effects differ substantially from the mean. We then calculate
\[
p_k^{\mathrm{CATE}}
=
\frac{
1+\sum_{b=1}^{B}
\mathbf{1}\{
Q_k^{(b)}\geq Q_k
\}
}{
B+1
}.
\]
We additionally perform a pooled test of treatment-effect heterogeneity across visits using the average of the visit-specific fourth-order statistics, $Q_{\mathrm{avg}}=m^{-1}\sum_{k=1}^{m}Q_k$.

\subsection{Comparators}

We selected SCOUT's ridge parameter $\eta$ by five-fold cross-validation over 12 logarithmically spaced values from $10^{-5}$ to $10^{1}$. We compared SCOUT with three alternative approaches to outcome learning and causal adjustment:

\begin{enumerate}[leftmargin=1.25em,label=(\alph*),labelsep=0.4em,itemsep=0.25em]
\item \textbf{Stable Balancing Weights with SCORE (SBW-SCORE)} \citep{zubizarreta2015stable,strobl2025score}: estimates nonnegative subject weights that minimize weight dispersion subject to approximate balance of the baseline covariates between treatment groups, then applies SCORE to the weighted longitudinal outcomes to learn a common treatment-responsive composite outcome. This serves as our covariate-balancing comparator.

\item \textbf{Inverse regression} \citep{zhang2025inverse}: regresses treatment assignment on the multivariate outcomes, uses the resulting coefficients to define a composite outcome that maximizes standardized treatment separation, and incorporates propensity weighting in observational settings to adjust for baseline confounding. This serves as our parsimonious linear comparator.

\item \textbf{Cross-validated targeted maximum likelihood estimation (CV-TMLE)} \citep{zhou2026cvtmle}: uses cross-validation to learn composite-outcome weights by optimizing a targeted treatment-effect estimate while estimating the treatment and outcome nuisance functions flexibly, after which the learned weights are fixed and TMLE is applied separately at each visit to estimate the ATE and CATE. This serves as our semiparametric causal-learning comparator.

\end{enumerate}

We additionally evaluated two ablations designed to isolate the contribution of SCOUT's low-dimensional adjustment strategy. \textbf{SCOUT-no-2D} omits the final reduction to two-dimensional adjustment scores, allowing us to assess the contribution of composite-specific dimensionality reduction to treatment-effect estimation. \textbf{SCOUT-no-COMPACT} replaces COMPACT's low-dimensional proxy representation with ridge-regularized adjustment using the observed baseline covariates while retaining SCOUT's outcome-learning framework. This tests whether explicitly recovering a low-dimensional adjustment representation provides an advantage over direct regularized adjustment in the original high-dimensional covariate space.

\subsection{Evaluation metrics}

We first assess \textbf{ATE error} using the absolute difference between the estimated ATE and the true sample-average treatment effect $\bar\tau_k^{\mathrm{true}}$ of the learned phenotype among individuals in $\mathcal I_k$, averaged across visits:
\[
\operatorname{ATE\ Error}
=
\frac{1}{m}
\sum_{k=1}^{m}
\left|
\widehat{\bar\tau}_k-\bar\tau_k^{\mathrm{true}}
\right|.
\]
Second, \textbf{CATE RMSE} measures subject-level recovery of the treatment-effect surface. Let $\tau_{ik}^{\mathrm{true}}$ denote the true CATE corresponding to the learned phenotype and the method's conditioning variables:
\[
\operatorname{CATE\ RMSE}
=
\frac{1}{m}
\sum_{k=1}^{m}
\left[
\frac{1}{n_k}
\sum_{i\in\mathcal I_k}
\left(
\widehat\tau_{ik}-\tau_{ik}^{\mathrm{true}}
\right)^2
\right]^{1/2}.
\]
CATE RMSE therefore evaluates how accurately each complete method estimates its intended CATE, rather than a CATE based on a common conditioning set. Imposing a shared conditioning set would either replace SCOUT's learned adjustment set, thereby nullifying a key methodological contribution, or require the competing algorithms to adopt SCOUT's adjustment set, effectively reducing them to minor modifications of SCOUT. The CATE conditioned on the learned, two-dimensional adjustment scores is an intentional and distinctive estimand of SCOUT.

Third, we quantify the \textbf{proportion of treatment-effect heterogeneity retained} by SCOUT's low-dimensional adjustment scores. Specifically, we compute 
\[
R_k^2
=
\frac{
\operatorname{Var}\left(
\operatorname{Proj}_{\bm S_k}
[\tau_k(\bm X)]
\right)
}{
\operatorname{Var}\left(\tau_k(\bm X)\right)
},
\]
which quantifies how much of the original treatment-effect heterogeneity is preserved by the low-dimensional representation. Lower values are preferable when treatment-effect estimation accuracy is maintained or improved, indicating that more of the original heterogeneity has been discarded without compromising estimation accuracy.

Fourth, we evaluate statistical calibration and power of the CRTs. \textbf{Type I error} is estimated as the proportion of null simulations yielding $p\leq 0.05$, considering both zero treatment effects and homogeneous nonzero treatment effects. We additionally assess \textbf{calibration across nominal significance thresholds} and \textbf{statistical power}, defined as the proportion of simulated alternatives with nonzero average or heterogeneous treatment effects that are correctly rejected.

Finally, we assess computational efficiency using the total wall-clock \textbf{runtime} required to complete each analysis.

\subsection{Synthetic Data Generation}

Simulation details are provided in Supplementary Material \ref{supp:synth_gen}. Briefly, we generated longitudinal data with latent patient characteristics that influenced both treatment assignment and multiple correlated outcomes, creating confounding that could only be addressed indirectly through noisy baseline proxies. Treatment effects varied across outcomes and visits, with patient-level heterogeneity determined by a combination of observed baseline characteristics. We systematically varied sample size and the number of observed proxies to evaluate estimation accuracy under different levels of available information. Separate CRT simulations evaluated Type I error for the ATE and CATE conditional independence tests using data generated with zero and constant additive treatment effects, respectively, and assessed statistical power under nonzero constant and heterogeneous treatment effects.

\section{Results}

\subsection{Synthetic Data}

\begin{figure}[t]
    \centering
    \includegraphics[width=0.85\linewidth]{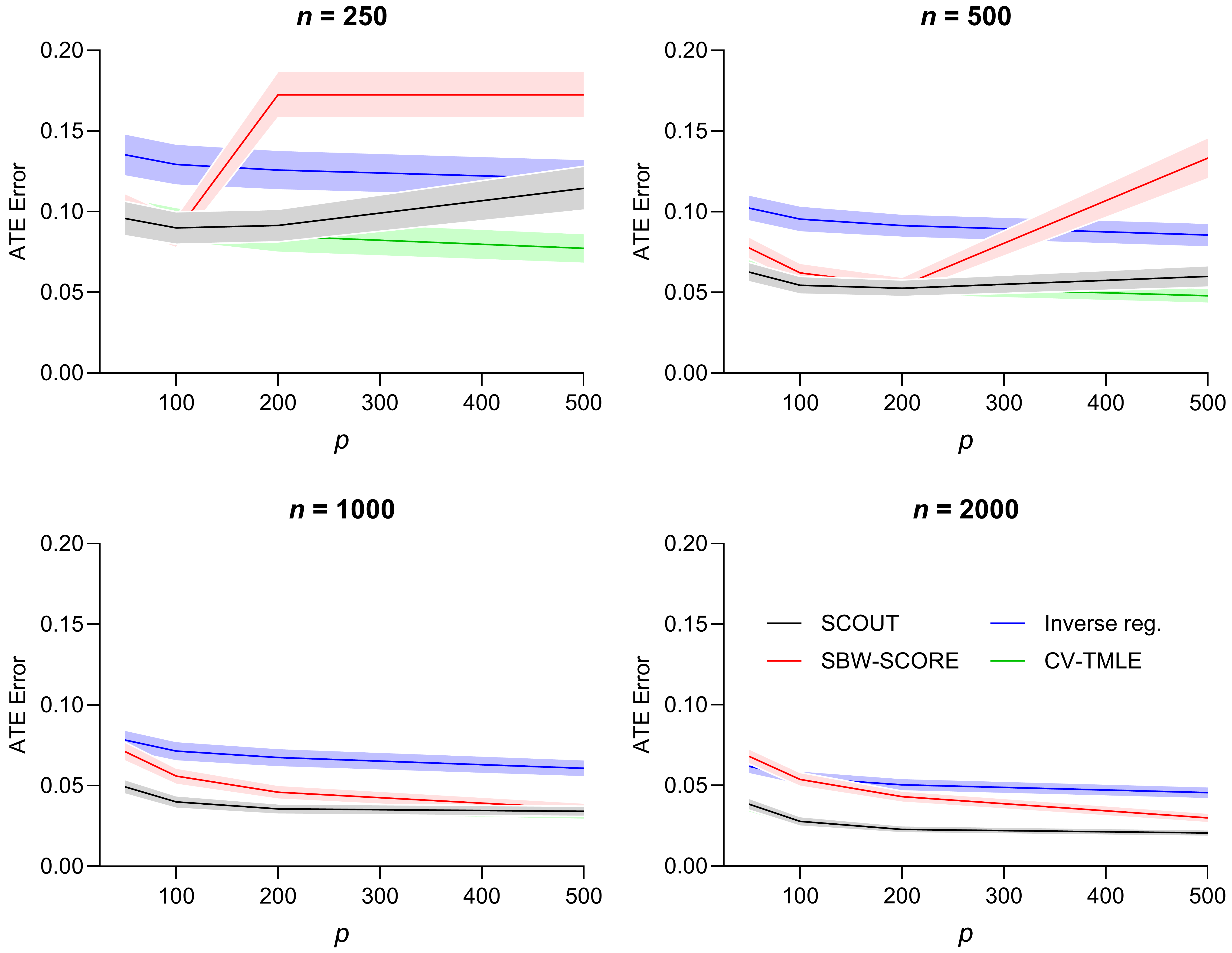}
    \caption{\textbf{ATE estimation error across simulation settings.} Mean ATE error is shown as a function of the number of baseline variables $p$, stratified by sample size $n$. Shaded regions denote 95\% confidence intervals across Monte Carlo replications. All panels share the same legend. In some panels, the CV-TMLE curve is not visible due to the overlying SCOUT curve.}
    \label{fig:sim:ATE}
\end{figure}

We generated synthetic observational data with latent confounding, noisy high-dimensional proxies, repeated multivariate outcomes, and heterogeneous treatment effects. We first compared algorithms using ATE error. CV-TMLE achieved the lowest error in the most challenging low-$n$, high-$p$ settings, although this advantage diminished rapidly as sample size increased and SCOUT converged to almost exactly the same performance (Figure \ref{fig:sim:ATE}). SCOUT generally outperformed the remaining methods across the evaluated settings. Ablation analyses further indicated that both COMPACT-based adjustment and the final two-dimensional refinement contributed to SCOUT's favorable ATE accuracy (Supplementary Figure \ref{fig:sim:ablations:ATE}).

\begin{figure}[t]
    \centering
    \includegraphics[width=0.85\linewidth]{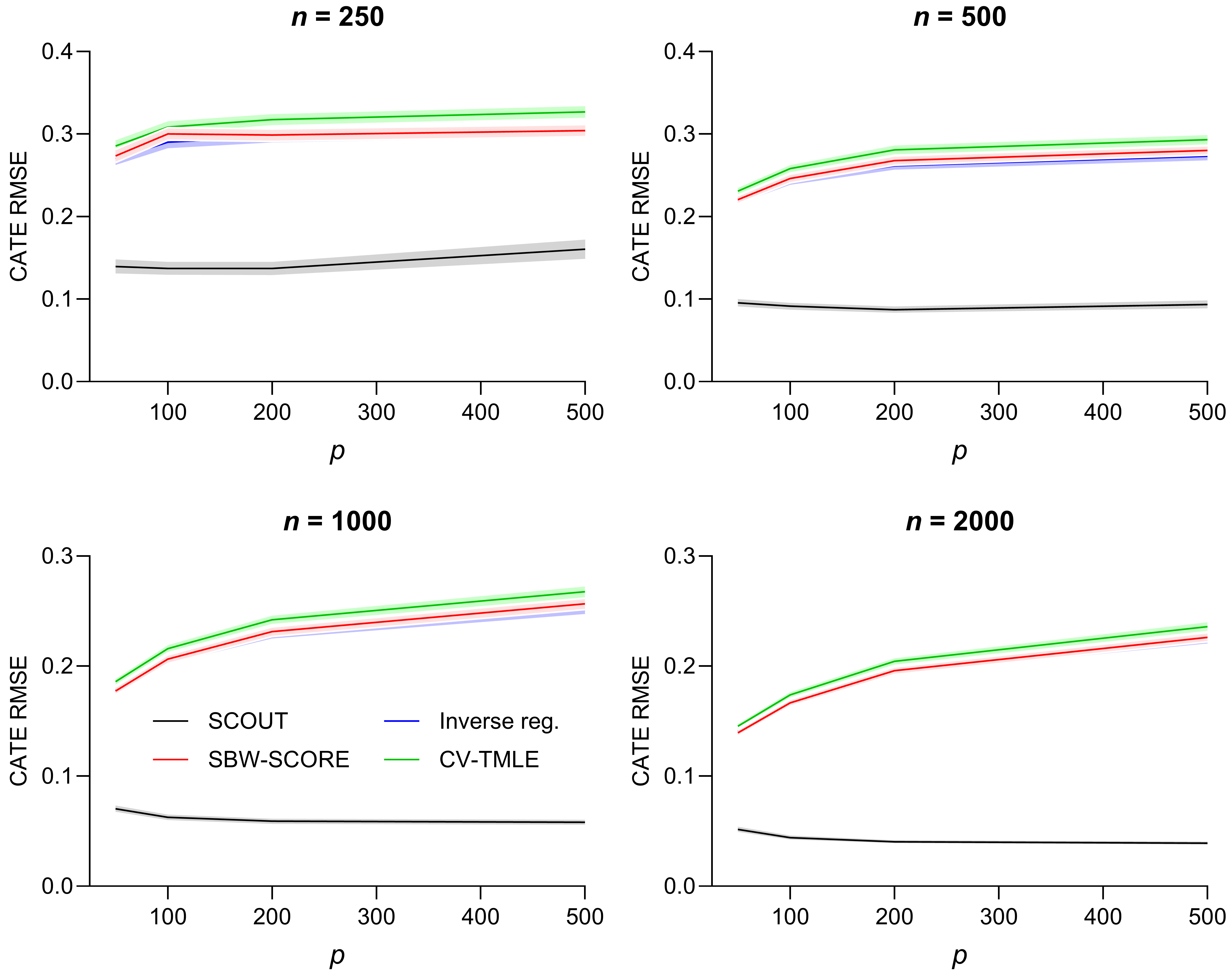}
    \caption{\textbf{CATE estimation error across simulation settings.} Mean CATE RMSE is shown as a function of $p$ for each sample size $n$, with error calculated relative to each method's corresponding CATE estimand.}
    \label{fig:sim:CATE}
\end{figure}

ATE error, however, is a coarse measure of estimation accuracy because individual-level treatment-effect errors can cancel when averaged. We therefore also evaluated CATE RMSE. SCOUT achieved substantially lower CATE error across all sample sizes and values of $p$ for its score-indexed CATE target (Figure \ref{fig:sim:CATE}). In contrast, CV-TMLE exhibited considerably larger CATE error despite accurately estimating the ATE, demonstrating that accurate recovery of the marginal treatment effect does not necessarily imply accurate recovery of heterogeneous treatment effects. Ablation analyses again showed that both COMPACT and the two-dimensional refinement contributed to minimizing CATE RMSE (Figure \ref{fig:sim:ablations:CATE}). SCOUT was therefore distinguished by simultaneously providing accurate ATE estimation and highly accurate estimation of the lower-dimensional CATE defined by its learned adjustment scores.

Importantly, SCOUT does not attempt to recover the full CATE conditional on all of $\bm X$. Instead, it targets the component of the $\bm X$-indexed CATE retained after projecting treatment-effect heterogeneity onto the learned adjustment scores $\bm S$. Consistent with this design, the score-indexed CATE retained less than 50\% of the variance of the full $\bm X$-indexed CATE in most settings (Figure \ref{fig:sim:R2}). The retained fraction decreased as $p$ increased, reaching approximately 25\% at $p=500$. This pattern illustrates a deliberate resolution--estimability tradeoff: as the dimensionality of $\bm X$ increases, SCOUT continues to represent treatment-effect heterogeneity through a low-dimensional score space rather than allowing the complexity of the CATE target to grow with $p$. The resulting CATE therefore captures only the component of $\bm X$-indexed heterogeneity represented by $\bm S$, but this lower-dimensional target can be estimated substantially more accurately.

SCOUT was computationally efficient across the simulation grid. Runtime remained below 1.2 seconds even at $p=500$, despite the algorithm's cubic dependence on $p$. In contrast, CV-TMLE was typically two to three orders of magnitude slower (Supplementary Figure \ref{fig:sim:time}), indicating that SCOUT's gains in heterogeneous-effect estimation were achieved without a substantial computational penalty.

Finally, the 95\% confidence intervals for empirical Type I error rates included the nominal 5\% level across all null scenarios (Supplementary Figure \ref{fig:sim:typeIerror_power}a). The CRT also demonstrated high statistical power across all simulated alternatives (Supplementary Figure \ref{fig:sim:typeIerror_power}b). Calibration curves showed close agreement between nominal significance levels and empirical rejection rates under the zero-ATE null, with good calibration under the homogeneous-CATE null at nominal significance levels up to 15\% (Supplementary Figure \ref{fig:sim:calibration}). Together, these results support the CRT as a reliable approach to statistical inference.

\subsection{Semi-Synthetic Data}

We next evaluated SCOUT using semi-synthetic data calibrated to the Clinical Antipsychotic Trials of Intervention Effectiveness (CATIE) study\cite{lieberman2005catie}. We estimated proxy-loading directions from 364 processed baseline features derived from CATIE and incorporated them into the synthetic data-generating mechanism, preserving the remaining causal structure and known treatment effects. The simulations used $n=1448$ individuals and varied the number of baseline proxies $p\in\{50,100,200,364\}$.

\begin{figure}[!t]
    \centering
    \includegraphics[width=1\linewidth]{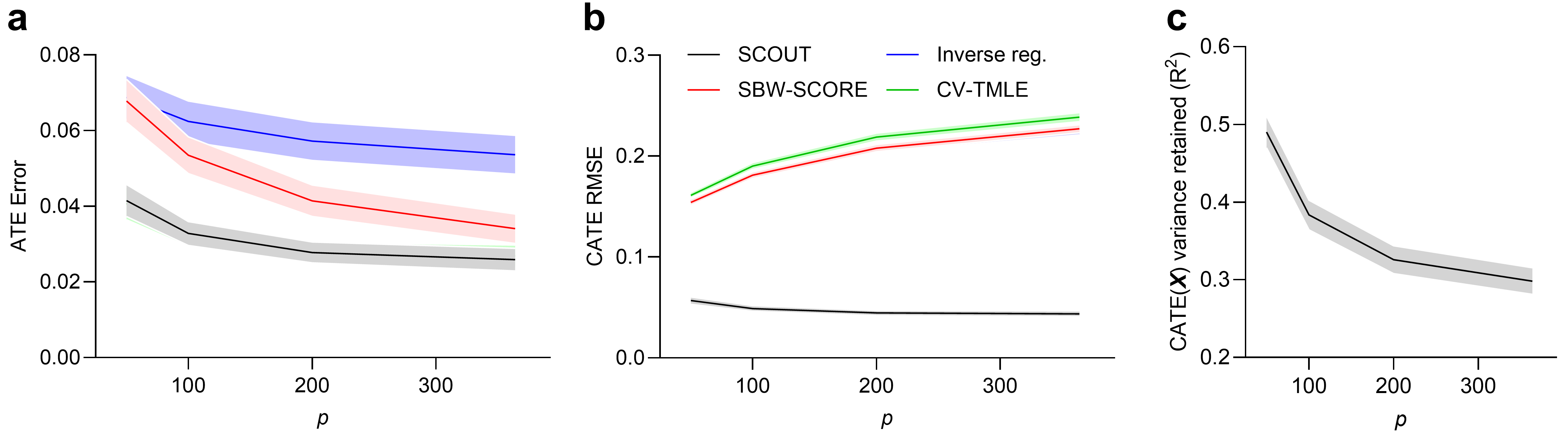}
    \caption{\textbf{Causal effect estimation in CATIE-based
semi-synthetic data.} (a) ATE error, (b) CATE RMSE,
and (c) proportion of CATE variance retained by SCOUT's
adjustment scores across baseline proxy dimensions.
CATE error is evaluated relative to each method's
corresponding estimand.}
    \label{fig:CATIE}
\end{figure}

SCOUT and CV-TMLE achieved nearly identical ATE error across all values of $p$ (Figure \ref{fig:CATIE}a). SCOUT also achieved substantially lower CATE RMSE for its score-indexed target than the competing methods achieved for their respective targets (Figure \ref{fig:CATIE}b). Both ablations generally exhibited higher ATE and CATE RMSE, with the advantage of SCOUT's two-dimensional refinement becoming more pronounced as $p$ increased (Supplementary Figure \ref{fig:CATIE:ablations}). As in the synthetic experiments, the score-indexed CATE retained only a fraction of the variance of the full $\bm X$-indexed CATE, decreasing from approximately 49\% to 30\% as $p$ increased (Figure \ref{fig:CATIE}c). SCOUT completed each analysis in less than 0.81 seconds, compared with up to 559 seconds for CV-TMLE (Supplementary Figure \ref{fig:CATIE:time}). Overall, the CATIE-based experiments closely reproduced the synthetic findings, demonstrating that SCOUT's competitive ATE accuracy and advantages in estimating score-indexed treatment heterogeneity persist under empirically calibrated proxy-loading structures.

We next evaluated the CRT's operating characteristics using 500 Monte Carlo replicates with $p=364$ baseline proxies. Each replicate again included zero, constant, and heterogeneous treatment-effect scenarios, with outcome-specific constant effects independently sampled from $\operatorname{Unif}(-1,1)$ and 500 conditional treatment randomizations performed per scenario. The CRT controlled Type I error at the nominal 5\% level for both ATE and CATE tests (Supplementary Figure \ref{fig:sim:catie_typeIerror_power}a). Calibration curves further showed close agreement between nominal significance levels and empirical rejection rates (Supplementary Figure \ref{fig:sim:catie_calibration}). The CRT also achieved high statistical power under all simulated alternatives (Supplementary Figure \ref{fig:sim:catie_typeIerror_power}b).

\subsection{Real Application}

We applied SCOUT to observational data from the Systematic Treatment Enhancement Program for Bipolar Disorder (STEP-BD). We sought to determine whether lithium-containing regimens exhibit distinct adverse-effect profiles compared with standard care without lithium, focusing on adverse effects observed while patients remained on their respective treatment strategies during the naturalistic treatment of bipolar disorder.

We defined treatment at the index visit as lithium-containing standard care ($T=1$) versus standard care without lithium ($T=0$). We identified lithium initiators at their first lithium-positive visit following a lithium-negative assessment within the preceding 60 days. We matched each initiator to up to three non-lithium controls with comparable pretreatment clinical characteristics and index visits within 15 study days.

\begin{figure}[!t]
    \centering
    \includegraphics[width=0.9\linewidth]{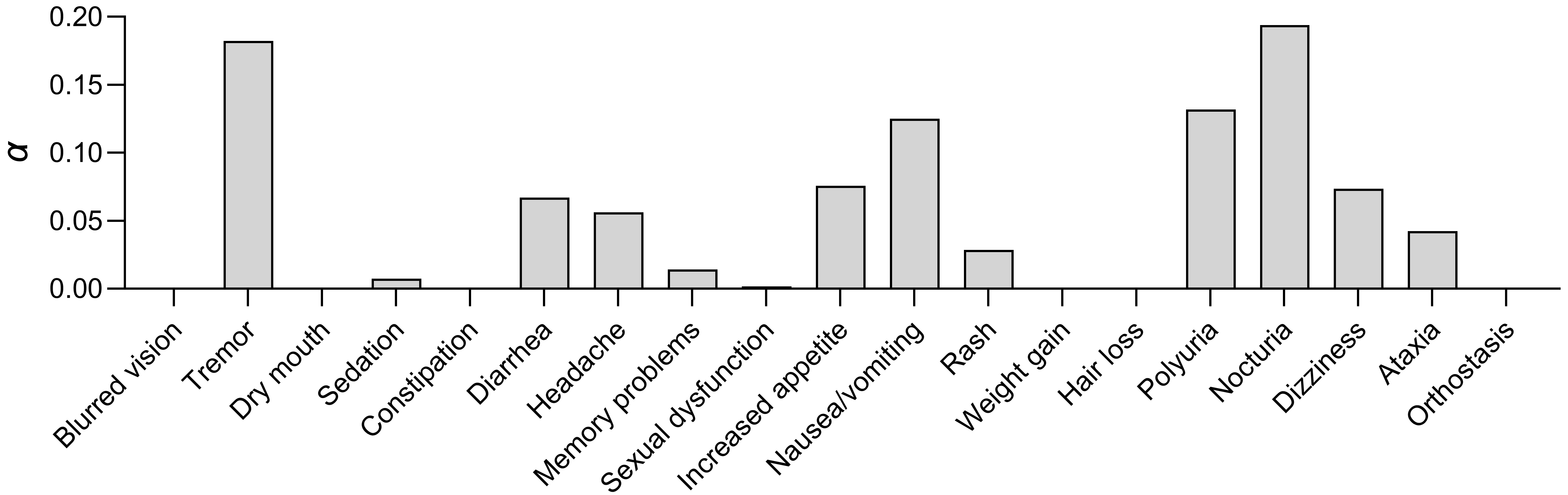}
    \caption{\textbf{Discovered adverse-effect composite.} SCOUT identified a nonnegatively weighted composite comprising multiple symptoms associated with lithium treatment. The weights indicate the relative contributions of individual symptoms to the discovered composite.}
    \label{fig:res:lithium_composite}
\end{figure}

We considered outcomes at 45, 60, 75, and 90 days after the index visit. The first outcome assessment was selected as the closest complete assessment within 15--60 days after index. At subsequent follow-ups, we used the last complete assessment obtained while the patient remained on the original treatment strategy, allowing assessments through days 75, 90, and 105, respectively. When no additional eligible assessment was available, the last observed on-strategy adverse-effect measurements were carried forward. Thus, follow-up outcomes represented the last observed adverse-effect state under the original treatment strategy.

We considered 19 adverse effects as candidate outcomes (Figure \ref{fig:res:lithium_composite} x-axis). All outcomes were standardized using means and standard deviations pooled across the four follow-up periods. Adjustment covariates included 102 variables spanning demographic characteristics, body weight, depressive and manic symptoms, baseline adverse-effect severities, clinical severity, functioning, concomitant medications, clinical pathway, study site, and treatment timing. All covariates were measured at the last medication-documented lithium-negative assessment preceding the index visit. We set the primary endpoints to the ATE averaged across the four follow-up visits and treatment-effect heterogeneity assessed using the average of the visit-specific fourth-order statistics.

The preprocessed dataset comprised 481 STEP-BD participants, including 145 lithium initiators and 336 non-lithium controls. SCOUT learned two-dimensional confounder adjustment scores at each follow-up visit, with substantial treatment overlap in the resulting score spaces (Supplementary Figure \ref{fig:res:STEPBD}a, c; Supplementary Figure \ref{fig:res:STEPBD2}e, g). Moreover, SCOUT identified a lithium-associated adverse-effect composite with consistently positive adjusted treatment effects across all four follow-up periods (mean ATE = 0.250; Holm-adjusted $p=0.0026$, based on 10,000 randomizations). The learned composite assigned its largest normalized weights to nocturia (19.4\%), tremor (18.2\%), polyuria (13.2\%), and nausea/vomiting (12.5\%), with smaller contributions from other adverse effects (Figure \ref{fig:res:lithium_composite}). Visualization of the CATEs over the learned score spaces revealed localized regions of elevated treatment effects, suggesting that certain patient subgroups may be particularly susceptible to lithium-associated adverse effects (Supplementary Figure \ref{fig:res:STEPBD}b, d; Supplementary Figure \ref{fig:res:STEPBD2}f, h). Consistent with this pattern, the constant-effect CRT detected significant treatment-effect heterogeneity in the across-visits analysis ($Q_{\mathrm{avg}} = 0.0249$, Holm-adjusted $p=0.033$).

These findings closely correspond to established clinical experience with lithium. Without prespecifying its expected adverse-effect profile, SCOUT recovered multiple characteristic manifestations of lithium treatment, including urinary, neurological, and gastrointestinal adverse effects \cite{Gitlin2016}. Furthermore, the detected treatment-effect heterogeneity reflects the well-recognized variability in lithium tolerability, whereby some patients experience substantially greater adverse-effect burdens than others \cite{nieuwoudt2026lithium}. Thus, SCOUT recovered both a clinically recognizable pharmacological profile and interindividual variation in treatment response directly from observational data.

\section{Discussion}

We developed SCOUT to learn treatment-responsive composite outcomes while controlling for confounding throughout the learning process using low-dimensional representations of high-dimensional, noisy baseline variables. After learning the composite, SCOUT constructs a targeted two-dimensional adjustment representation at each follow-up visit to estimate ATEs and score-indexed CATEs. We also introduced CRTs that account for adaptive outcome learning and confounding adjustment, including a fourth-order statistic that emphasizes large deviations from the ATE. This statistic targets clinically important heterogeneity that arises when a small subgroup of patients experiences substantial treatment benefits or adverse effects.

A central feature of SCOUT is its estimation of CATEs conditional on the learned adjustment scores. As biomedical datasets grow, the number of potential effect modifiers can become extremely large, making accurate CATE estimation conditional on the full baseline covariate vector increasingly difficult. SCOUT instead restricts the conditioning space to two dimensions at each visit, using scores that summarize baseline information associated with treatment assignment and subsequent outcomes; these scores capture clinical characteristics that influence treatment decisions and prognosis. Our experiments showed that the scores retained only a fraction of the treatment-effect heterogeneity represented by the full covariate space. By discarding this additional information, SCOUT reduced CATE estimation to a more tractable problem in high-dimensional settings, enabling substantially lower estimation error for its score-indexed estimand.

Although SCOUT can learn an interpretable composite with a large treatment effect, this alone does not guarantee its clinical utility. We therefore recommend defining outcome panels around clinically important, coherent domains rather than indiscriminately including all available outcomes \cite{strobl2026taskaligned}. For example, a psychiatric application might focus on clinically important symptom severity dimensions within a specific disorder rather than combine outcomes spanning unrelated diagnoses. Such targeted panels help SCOUT learn clinically meaningful composites that capture the causal effects of the treatment under study. Our STEP-BD application illustrates this approach: we selected adverse effects plausibly relevant to distinguishing the tolerability profiles of lithium and antipsychotics, rather than including unrelated adverse events, allowing SCOUT to recover a recognizable lithium-associated profile.

Several opportunities for further methodological development remain. First, SCOUT currently relies on linear representations and treatment-effect models. Nonlinear extensions using basis functions or reproducing kernels could capture more complex confounding and treatment-response relationships. More flexible estimation of the treatment assignment mechanism may also improve CRT calibration. Second, the current CRT assumes a fixed treatment assignment across follow-up visits and does not accommodate time-varying treatment decisions. This restricted our STEP-BD analysis to outcomes observed under the initial treatment strategies, rather than allowing treatment changes to be modeled explicitly. Third, although SCOUT accommodates high-dimensional baseline variables, its initial adjustment representation contains up to $r+1$ dimensions per visit, where $r$ is the number of outcome components. As $r$ increases, this intermediate representation becomes less parsimonious, diminishing the computational and statistical advantages of dimensionality reduction. Future work should therefore investigate nonlinear extensions, more aggressive but principled reduction or regularization of the initial adjustment representation, and inference under time-varying treatment strategies.

\section{Conclusion}

SCOUT simultaneously addresses three very frequent challenges in biomedical observational research: confounding represented by numerous noisy proxies of latent variables, uncertainty about which outcomes capture treatment response, and the difficulty of estimating heterogeneous treatment effects in high dimensions. SCOUT addresses these challenges by constructing low-dimensional confounding adjustment scores, learning treatment-responsive composite outcomes, and estimating CATEs conditional on the learned scores rather than the full covariate space. Furthermore, SCOUT provides CRTs for ATEs and CATE heterogeneity that account for adaptive outcome learning and confounding adjustment. Together, these capabilities make SCOUT a promising tool for investigators studying causal treatment effects through target trial emulations using large observational datasets, such as electronic health records, longitudinal disease registries, and prospective observational cohorts.

\section*{Code Availability}

An R implementation of SCOUT is available at \url{https://github.com/ericstrobl/SCOUT}.

\bibliographystyle{elsarticle-num}
\bibliography{biblio}

\section{Supplementary Material}

\setcounter{figure}{0}
\renewcommand{\thefigure}{S\arabic{figure}}
\renewcommand{\figurename}{Supplementary Figure}

\subsection{Pseudocode} \label{supp:sec:pseudocode}

We summarize the complete SCOUT procedure in
Algorithm~\ref{alg:SCOUT}. The algorithm takes as input a
high-dimensional baseline proxy matrix, a binary treatment
vector, and repeated multivariate outcome measurements.
At each visit, SCOUT first computes the treatment-residualized
outcome-direction matrix $\mathbf{V}_{R,k}$
(Line~\ref{alg:SCOUT:VR}) and constructs the initial
adjustment scores using the treatment direction and all
$r$ outcome directions
(Lines~\ref{alg:SCOUT:universal}--\ref{alg:SCOUT:initial_scores}).
These scores provide a common adjustment representation
for all candidate outcome composites. SCOUT then
residualizes the complete outcome panel and treatment
with respect to these scores
(Line~\ref{alg:SCOUT:residualize}), removing the variation
explained by the initial adjustment representation.

Using the residualized outcomes and treatment, SCOUT
learns a single nonnegative outcome weight vector
$\widehat{\bm\alpha}$ shared across all visits by solving
the NNLS problem in \eqref{eq:NNLS}
(Line~\ref{alg:SCOUT:alpha}). Each visit contributes
equally to the objective through normalization by its
sample size $n_k$, preventing visits with more observations
from dominating outcome learning. Because the initial
adjustment scores are fixed throughout this optimization,
SCOUT can evaluate different outcome combinations without
recomputing the adjustment representation. When at least
one learned weight is nonzero, the weights are subsequently
normalized to sum to one for interpretability
(Line~\ref{alg:SCOUT:normalize}).

Once the outcome weights have been learned, SCOUT returns
to the original, unresidualized clinical outcomes to
construct the composite $\mathbf Y_k^*$ at each visit
(Line~\ref{alg:SCOUT:phenotype}). It then replaces the
initial $(r+1)$-dimensional adjustment
representation with a two-dimensional
representation tailored to the discovered phenotype
(Lines~\ref{alg:SCOUT:scalar_compact}--\ref{alg:SCOUT:final_scores}). Finally, SCOUT fits
the treatment-effect model in \eqref{eq:causal_model}
using the refined adjustment scores and computes
patient-specific CATEs and the empirical ATE at each
visit (Lines~\ref{alg:SCOUT:effect_model}--\ref{alg:SCOUT:effects}). The same outcome composite
is therefore used across all visits, while the
adjustment scores and estimated treatment effects
remain visit-specific.

\begin{algorithm}[H]
\caption{SCOUT: Score-based Causal Outcome learning Using Treatment signals}
\label{alg:SCOUT}
\begin{algorithmic}[1]

\Statex \textbf{Input:}
Baseline proxy matrix $\mathbf{X}\in\mathbb R^{n\times p}$;
binary treatment vector $\mathbf{T}\in\{0,1\}^n$;
visit-specific outcome matrices
$\{\mathbf{Y}_k\}_{k=1}^{m}$;
observed individual sets $\{\mathcal I_k\}_{k=1}^{m}$;
ridge parameter $\eta$

\Statex \textbf{Output:}
Learned outcome weights $\widehat{\bm\alpha}$;
adjustment scores $\{\mathbf{S}_k\}_{k=1}^{m}$;
visit-specific ATEs $\{\widehat{\bar\tau}_k\}_{k=1}^{m}$
and CATEs $\{\widehat\tau_{ik}\}$

\For{$k=1,\ldots,m$}

    \State $\mathbf{X}_k\leftarrow\mathbf{X}[\mathcal I_k,:]$,
    $\mathbf{T}_k\leftarrow\mathbf{T}[\mathcal I_k]$,
    $n_k\leftarrow|\mathcal I_k|$

    \State $\mathbf{H}_{n_k}\leftarrow
    \mathbf{I}_{n_k}
    -n_k^{-1}\bm 1\bm 1^\top$

    \State $\mathbf{X}_{c,k}\leftarrow
    \mathbf{H}_{n_k}\mathbf{X}_k$,
    $\mathbf{T}_{c,k}\leftarrow
    \mathbf{H}_{n_k}\mathbf{T}_k$

    \State $\mathbf{M}_{T,k}\leftarrow
    \mathbf{H}_{n_k}
    -\mathbf{T}_{c,k}
    (\mathbf{T}_{c,k}^\top\mathbf{T}_{c,k})^{-1}
    \mathbf{T}_{c,k}^\top$

    \State $\bm v_{T,k}\leftarrow
    n_k^{-1}\mathbf{X}_{c,k}^\top\mathbf{T}_{c,k}$

    \State $\displaystyle
    \mathbf{G}_{k,\eta}\leftarrow
    \frac{1}{2n_k}
    \left(
    \mathbf{X}_{c,k}^\top\mathbf{X}_{c,k}
    +
    \mathbf{X}_{c,k}^\top
    \mathbf{M}_{T,k}\mathbf{X}_{c,k}
    \right)+\eta\mathbf{I}_p$

    \State $\mathbf{V}_{R,k}\leftarrow
    n_k^{-1}\mathbf{X}_{c,k}^\top
    \mathbf{M}_{T,k}\mathbf{Y}_k$
    \label{alg:SCOUT:VR}

    \State $\displaystyle
    \mathbf{D}_k^{(0)}\leftarrow
    \left[
    \mathbf{G}_{k,\eta}^{-1}\bm v_{T,k},\;
    \mathbf{G}_{k,\eta}^{-1}\mathbf{V}_{R,k}
    \right]$
    \label{alg:SCOUT:universal}

    \State $\mathbf{S}_k^{(0)}
    \leftarrow\mathbf{X}\mathbf{D}_k^{(0)}$
    \label{alg:SCOUT:initial_scores}

    \State Residualize $\mathbf{Y}_k$ and $\mathbf{T}_k$
    on $[\bm 1,\mathbf{S}_k^{(0)}[\mathcal I_k,:]]$ to obtain
    $\widetilde{\mathbf{Y}}_k$ and
    $\widetilde{\mathbf{T}}_k$
    \label{alg:SCOUT:residualize}

\EndFor

\State Solve the NNLS problem in \eqref{eq:NNLS}
for $\widehat{\bm\alpha}\geq\bm 0$,
weighting each visit by $1/(mn_k)$
\label{alg:SCOUT:alpha}

\If{$\sum_{j=1}^{r}\widehat{\alpha}_j>0$}
    \State $\displaystyle
    \widehat{\bm\alpha}\leftarrow
    \frac{\widehat{\bm\alpha}}
    {\sum_{j=1}^{r}\widehat{\alpha}_j}$
    \label{alg:SCOUT:normalize}
\EndIf

\For{$k=1,\ldots,m$}

    \State $\mathbf Y_k^{*}\leftarrow
    \mathbf{Y}_k\widehat{\bm\alpha}$
    \label{alg:SCOUT:phenotype}

    \State $\bm v_{R,k}^{*}\leftarrow
    \mathbf{V}_{R,k}\widehat{\bm\alpha}$
    \label{alg:SCOUT:scalar_compact}

    \State $\displaystyle
    \mathbf{D}_k\leftarrow
    \left[
    \mathbf{G}_{k,\eta}^{-1}\bm v_{T,k},\;
    \mathbf{G}_{k,\eta}^{-1}\bm v_{R,k}^{*}
    \right]$

    \State $\mathbf{S}_k\leftarrow\mathbf X\mathbf D_k$
    \label{alg:SCOUT:final_scores}

    \State Fit \eqref{eq:causal_model} by ordinary least
    squares using $(\mathbf Y_k^{*},\mathbf{T}_k,\mathbf{S}_k[\mathcal I_k,:])$
    \label{alg:SCOUT:effect_model}

    \State For each $i\in\mathcal I_k$, compute
    $\widehat\tau_{ik}\leftarrow
    \widehat\beta_{T,k}
    +\bm S_{k,i}^{\top}\widehat{\bm\beta}_{TS,k}$

    \State $\displaystyle
    \widehat{\bar\tau}_k\leftarrow
    \frac{1}{n_k}
    \sum_{i\in\mathcal I_k}\widehat\tau_{ik}$
    \label{alg:SCOUT:effects}

\EndFor

\end{algorithmic}
\end{algorithm}

\subsection{Computational complexity}
\label{app:complexity}

We analyze the computational complexity of Algorithm~\ref{alg:SCOUT}
for a fixed ridge parameter $\eta$. Because the individuals
observed at different visits may differ, SCOUT generally
constructs the adjustment matrices separately
at each visit. Forming the proxy Gram matrix
$\mathbf{X}_{c,k}^{\top}\mathbf{X}_{c,k}$ costs
$O(n_kp^2)$, while the treatment-adjusted covariance
matrix can be computed from this Gram matrix using a
rank-one correction at an additional cost of $O(p^2)$.
Eigendecomposing the resulting $p\times p$ matrix
used to construct $\mathbf{G}_{k,\eta}^{-1}$ costs
$O(p^3)$. Computing the treatment-residualized
outcome-direction matrix $\mathbf{V}_{R,k}$ in
Line~\ref{alg:SCOUT:VR} requires multiplication of
the $n_k\times p$ proxy matrix by the
$n_k\times r$ outcome matrix, costing $O(n_kpr)$.
Thus, preprocessing at visit $k$ costs
\[
O(n_kp^2+p^3+n_kpr).
\]

The initial adjustment representation contains at most
$r+1$ scores per visit. Applying $\mathbf{G}_{k,\eta}^{-1}$
to the treatment and outcome directions in
Line~\ref{alg:SCOUT:universal} costs $O(p^2r)$. Constructing the initial
scores in Line~\ref{alg:SCOUT:initial_scores} costs
$O(npr)$. Residualizing the $r$ outcomes and
treatment with respect to these scores, together
with forming the residualized outcome cross-products,
costs $O(n_kr^2)$. Consequently, the initial adjustment and residualization
stages across all visits require
\[
O\left(
\sum_{k=1}^{m}
\left[
n_kp^2+p^3+p^2r+npr+n_kr^2
\right]
\right).
\]

Once the residualized treatment and outcome matrices
have been obtained, SCOUT learns the outcome weights
by solving the NNLS problem in \eqref{eq:NNLS}.
Expanding the objective and omitting terms independent
of $\bm\alpha$ gives
\[
\widehat{\bm\alpha}
=
\arg\min_{\bm\alpha\geq\bm 0}
\left\{
\bm\alpha^\top\mathbf{H}\bm\alpha
-
2\bm c^\top\bm\alpha
\right\},
\]
where
\[
\mathbf{H}
=
\frac{1}{m}\sum_{k=1}^{m}
\frac{1}{n_k}
\widetilde{\mathbf{Y}}_k^\top
\widetilde{\mathbf{Y}}_k,
\qquad
\bm c
=
\frac{1}{m}\sum_{k=1}^{m}
\frac{1}{n_k}
\widetilde{\mathbf{Y}}_k^\top
\widetilde{\mathbf{T}}_k.
\]
Computing these quantities costs $O(Nr^2)$,
which is already included in the preceding
residualization cost. Once they have been computed,
evaluating the NNLS objective or gradient requires
only operations on an $r\times r$ matrix and an
$r$-dimensional vector, costing $O(r^2)$ per
evaluation. If $N_{\mathrm{opt}}$ denotes the
number of objective and gradient evaluations,
the NNLS optimization contributes
$O(N_{\mathrm{opt}}r^2)$. Importantly, optimization
does not require further operations involving
the original $p$ baseline variables.

After learning $\widehat{\bm\alpha}$, SCOUT constructs
the scalar outcome
$\mathbf Y_k^*=\mathbf{Y}_k\widehat{\bm\alpha}$
at a cost of $O(n_kr)$ per visit.
The corresponding residual-outcome direction
can be obtained directly as
$\bm v_{R,k}^{*}
=\mathbf{V}_{R,k}\widehat{\bm\alpha}$,
costing $O(pr)$.
Because the treatment-dependent matrix
$\mathbf{G}_{k,\eta}$ has already been
decomposed, constructing the final two-dimensional
adjustment representation requires at most $O(p^2)$
additional operations per visit.
Computing the two final adjustment scores costs
$O(np)$, and fitting the treatment-effect
model in Line~\ref{alg:SCOUT:effect_model},
including treatment-by-score interactions,
costs $O(n_k)$.
The complete post-selection stage therefore costs
\[
O\left(
\sum_{k=1}^{m}
[p^2+np+n_kr]
\right),
\]
which is absorbed by the costs of the initial
adjustment stage.

Combining these terms, the total complexity of
SCOUT for fixed $\eta$ is
\[
O\left(
\sum_{k=1}^{m}
\left[
n_kp^2
+p^3
+p^2r
+npr
+n_kr^2
\right]
+
N_{\mathrm{opt}}r^2
\right).
\]
Equivalently, writing $N=\sum_{k=1}^{m}n_k$,
\[
O\left(
mp^3+Np^2
+mp^2r+mnpr+Nr^2
+N_{\mathrm{opt}}r^2
\right).
\]
When all $n$ individuals are observed at every visit,
the treatment-dependent Gram matrix and its
eigendecomposition can be shared across visits.
In this case, the complexity reduces to
\[
O\left(
p^3+np^2
+
m\left[
p^2r+npr+nr^2
\right]
+
N_{\mathrm{opt}}r^2
\right).
\]

\subsection{Details of Synthetic Data Generation} \label{supp:synth_gen}

Each simulation contained $q=20$ latent patient characteristics, $r=20$ outcomes measured at $m=3$ post-treatment visits, and up to $p=500$ observed baseline proxies. For each Monte Carlo replicate, we first generated a master dataset with $n_{\max}=2000$ subjects and $p_{\max}=500$ proxies and then formed the different $(n,p)$ settings by taking nested subsets of subjects and covariates.

For subject $i$, the latent clinical state was generated as $\bm C\sim N(\bm 0,\mathbf{I}_q)$. Treatment assignment followed $T\sim\operatorname{Bernoulli}(e)$ with $e=\operatorname{logit}^{-1}(\bm C^\top\bm a)$. The elements of $\bm a$ had randomly sampled signs and magnitudes drawn uniformly from $[0.25,1]$ and were scaled by $1/\sqrt q$. Treatment therefore depended on an unobserved latent state that also affected the outcomes.

The observed baseline variables were noisy linear proxies of the latent state. We generated a loading matrix $\bm\Lambda\in\mathbb R^{q\times p}$ with independent standard normal entries and unit-normalized columns, and then set $\bm X=\sqrt{0.5}\bm\Lambda^\top\bm C+\sqrt{0.5}\bm\xi$, where $\bm\xi\sim N(\bm 0,\mathbf{I}_p)$. Thus, approximately half of the variance in each proxy arose from the shared latent structure and half from independent noise.

Baseline outcome means were generated from the same latent state as $\bm C^\top\bm\Gamma$, where $\bm\Gamma\in\mathbb R^{q\times r}$ contains the latent-state effects on the $r$ outcomes. Its entries had randomly sampled signs and magnitudes in $[0.25,1]$ and were scaled by $1/\sqrt q$. The columns of $\bm\Gamma$ were oriented so that treatment assignment and baseline outcome levels were confounded in the same direction across outcomes, preventing confounding from canceling when nonnegative outcome composites were formed.

Residual outcome variation was correlated both across outcomes and across visits. Within-visit correlations followed an AR(1) structure with parameter $\rho_{\mathrm{item}}=0.50$, and correlations across visits followed an AR(1) structure with $\rho_{\mathrm{time}}=0.50$. The resulting joint error covariance was the  product of the visit- and outcome-level correlation matrices.

Treatment effects contained both marginal and heterogeneous components. Outcome-specific marginal effects $\delta_j$ had randomly sampled signs and magnitudes in $[0.25,1]$, with draws restricted to contain both positive and negative effects. Heterogeneity was driven by a standardized linear combination $M$ of the first 10 observed proxies, with its direction chosen to be associated with treatment assignment. Outcome-specific heterogeneity loadings $\zeta_j$ were positive and sampled from $\operatorname{Unif}(0.5,1.5)$ before normalization. Treatment-effect magnitudes increased across visits
according to $(s_1,s_2,s_3)=(0.50,0.75,1.00)$ and were generated as
\[
\tau_{ikj}
=
s_k\delta_j
+
0.50\,M s_k\zeta_j.
\]
Observed outcomes were then generated as $\bm Y_{ik}=\bm\Gamma^\top \bm C_i+\bm\varepsilon_{ik}+T_i\bm\tau_{ik}$. Null datasets used the same latent, proxy, treatment, and noise-generating mechanisms but set all treatment effects to zero.

We generated 100 independent Monte Carlo replicates and evaluated $n\in\{250,500,1000,2000\}$ and $p\in\{50,100,200,500\}$, yielding 16 sample-size and dimensionality settings per replicate.

The CRT simulations used the same data-generating framework but fixed the sample size at $n=1000$ and the number of observed proxies at $p=200$, with 500 Monte Carlo replicates and 500 conditional randomizations per replicate and condition. Unlike the estimation accuracy simulations, the CRT simulations retained the same underlying structural parameters across Monte Carlo replicates while independently generating new samples. Type I error was evaluated for the ATE conditional-independence test under zero treatment effects and for the CATE constant-shift conditional-independence test under nonzero constant treatment effects. For the latter, outcome-specific constant effects were independently sampled from $\operatorname{Unif}(-1,1)$ in each replicate and scaled across visits by $s_k$, allowing positive, negative, and arbitrarily small effects while remaining constant across individuals. Statistical power was evaluated under nonzero constant and heterogeneous treatment effects for the ATE and CATE tests, respectively.

\subsection{Extra Synthetic Results}

\begin{figure}[H]
    \centering
    \includegraphics[width=0.9\linewidth]{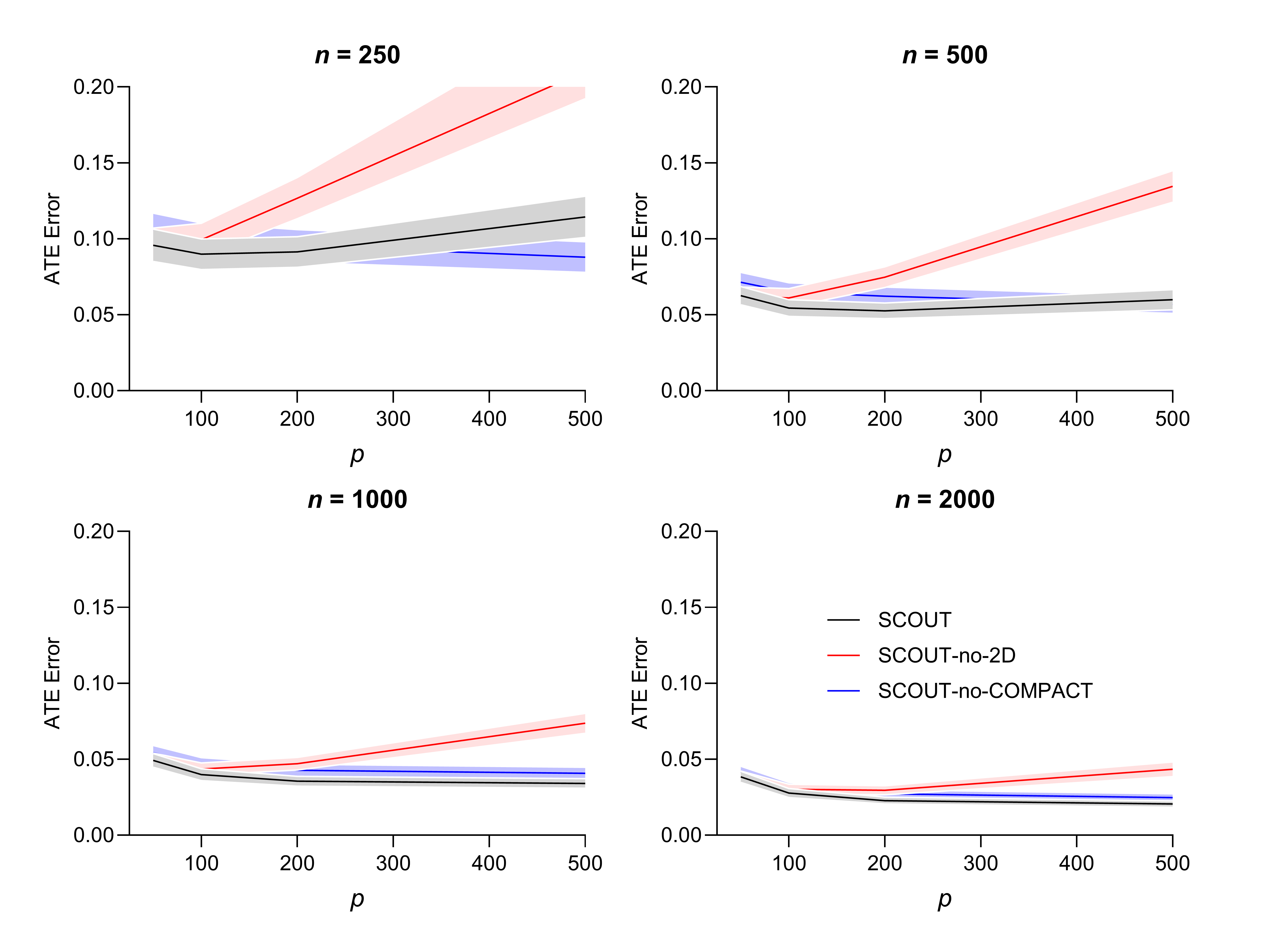}
    \caption{\textbf{Ablation analysis of ATE estimation.} Mean ATE error is shown across values of $p$ for SCOUT and variants omitting the two-dimensional refinement or COMPACT component.}
    \label{fig:sim:ablations:ATE}
\end{figure}

\begin{figure}[H]
    \centering
    \includegraphics[width=0.9\linewidth]{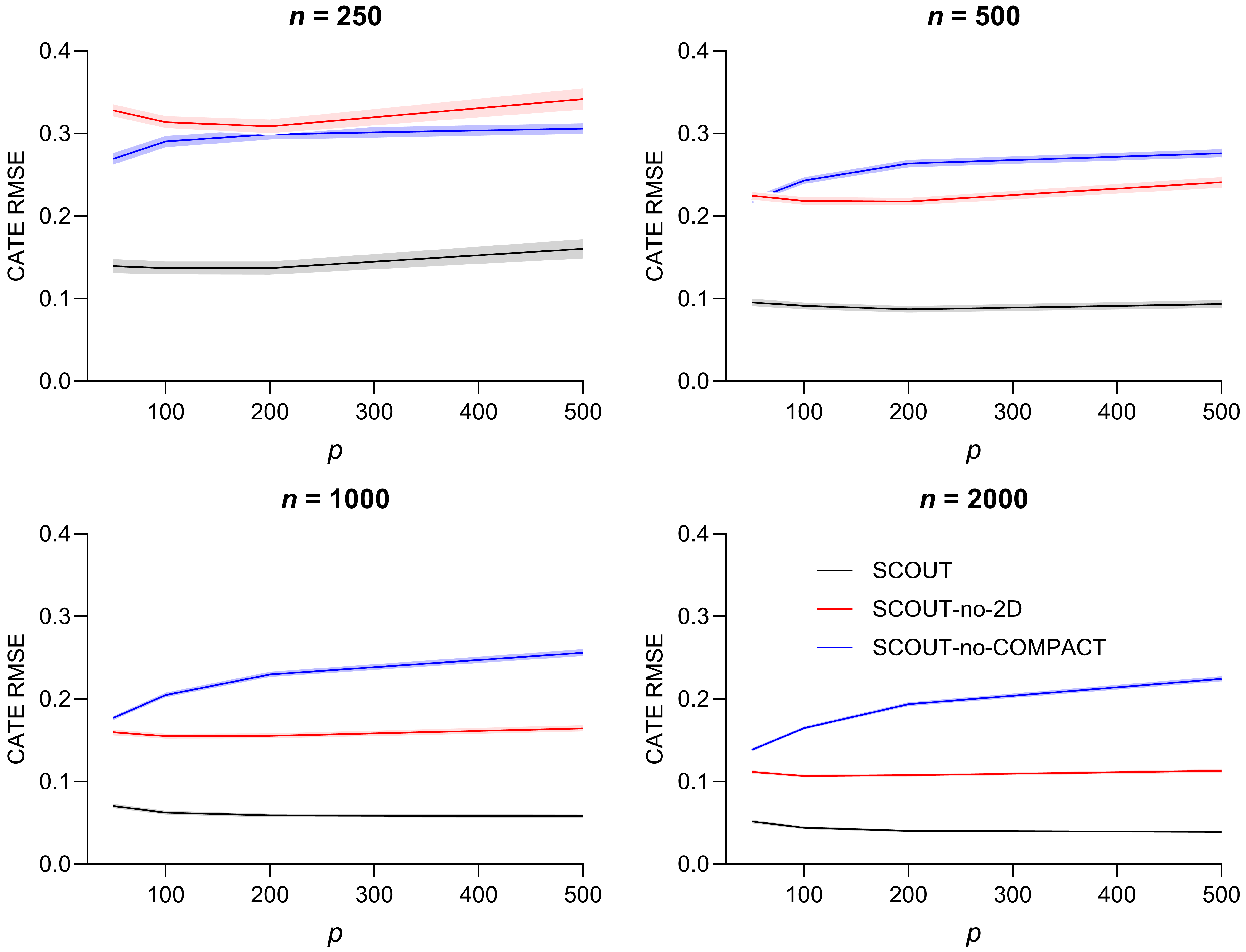}
    \caption{\textbf{Ablation analysis of CATE estimation.}}
    \label{fig:sim:ablations:CATE}
\end{figure}

\begin{figure}[H]
    \centering
    \includegraphics[width=0.45\linewidth]{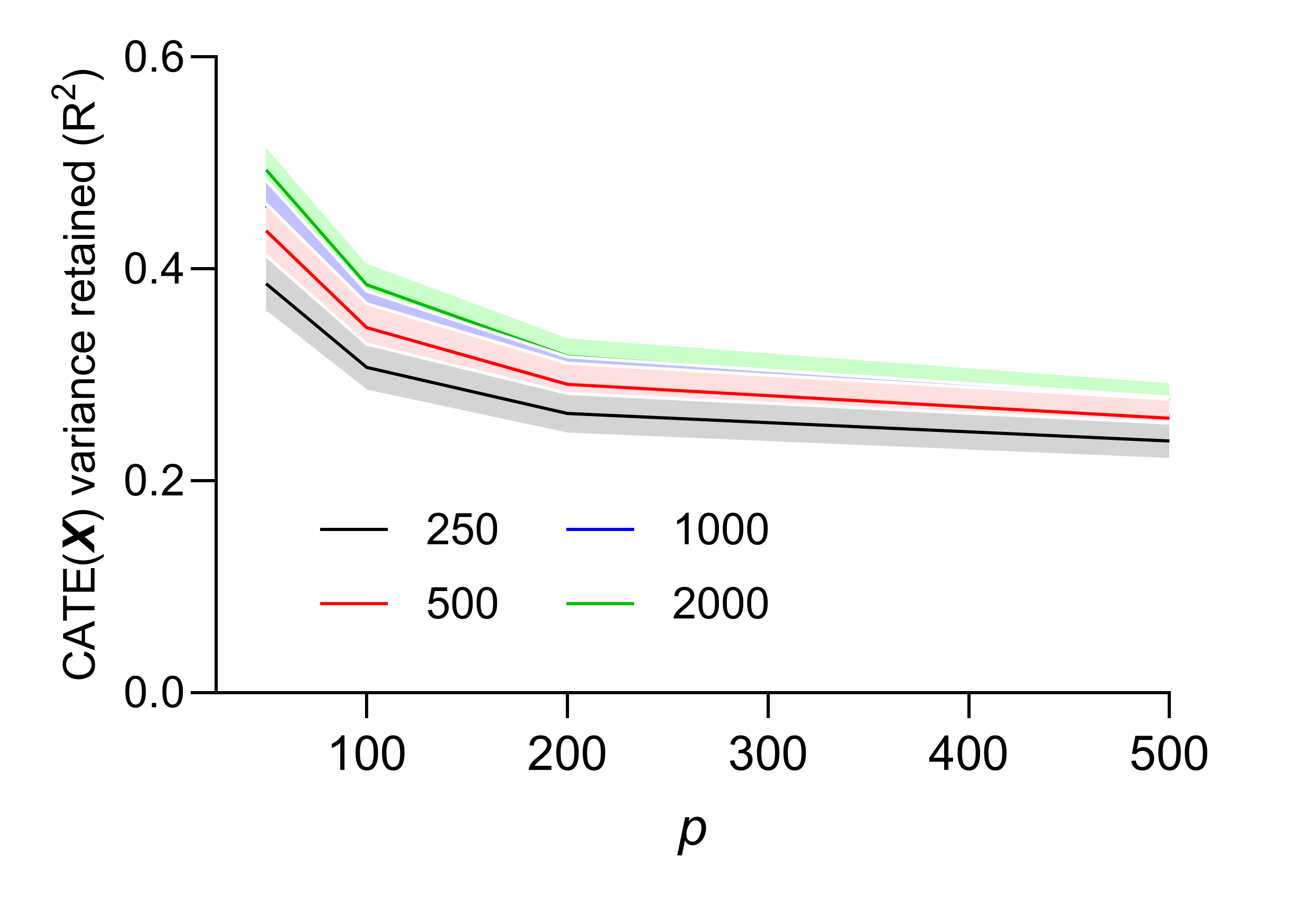}
    \caption{\textbf{Variance in the $\bm X$-indexed CATE retained by the SCOUT score representation.} $R^2$ denotes the fraction of variance in $\tau(\bm X)$ captured by its projection onto the learned adjustment scores $\bm S$. Lines correspond to sample size $n$.}
    \label{fig:sim:R2}
\end{figure}

\begin{figure}[H]
    \centering
    \includegraphics[width=0.9\linewidth]{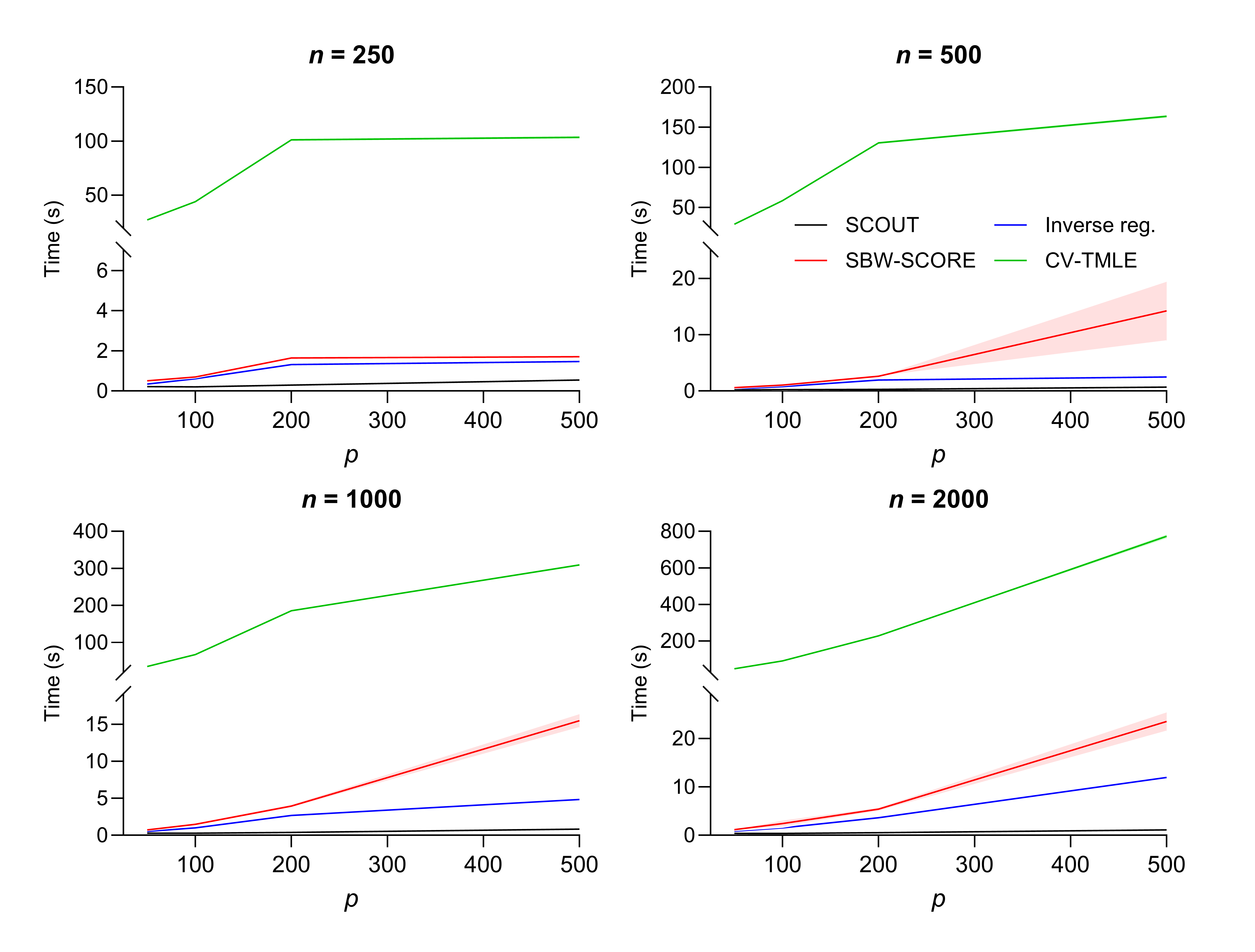}
    \caption{\textbf{Computational runtime across simulation settings.} Mean wall-clock time is shown as a function of $p$ for each sample size $n$. Error bands for some lines are not visible because they are too narrow.}
    \label{fig:sim:time}
\end{figure}

\begin{figure}[H]
    \centering
    \includegraphics[width=0.9\linewidth]{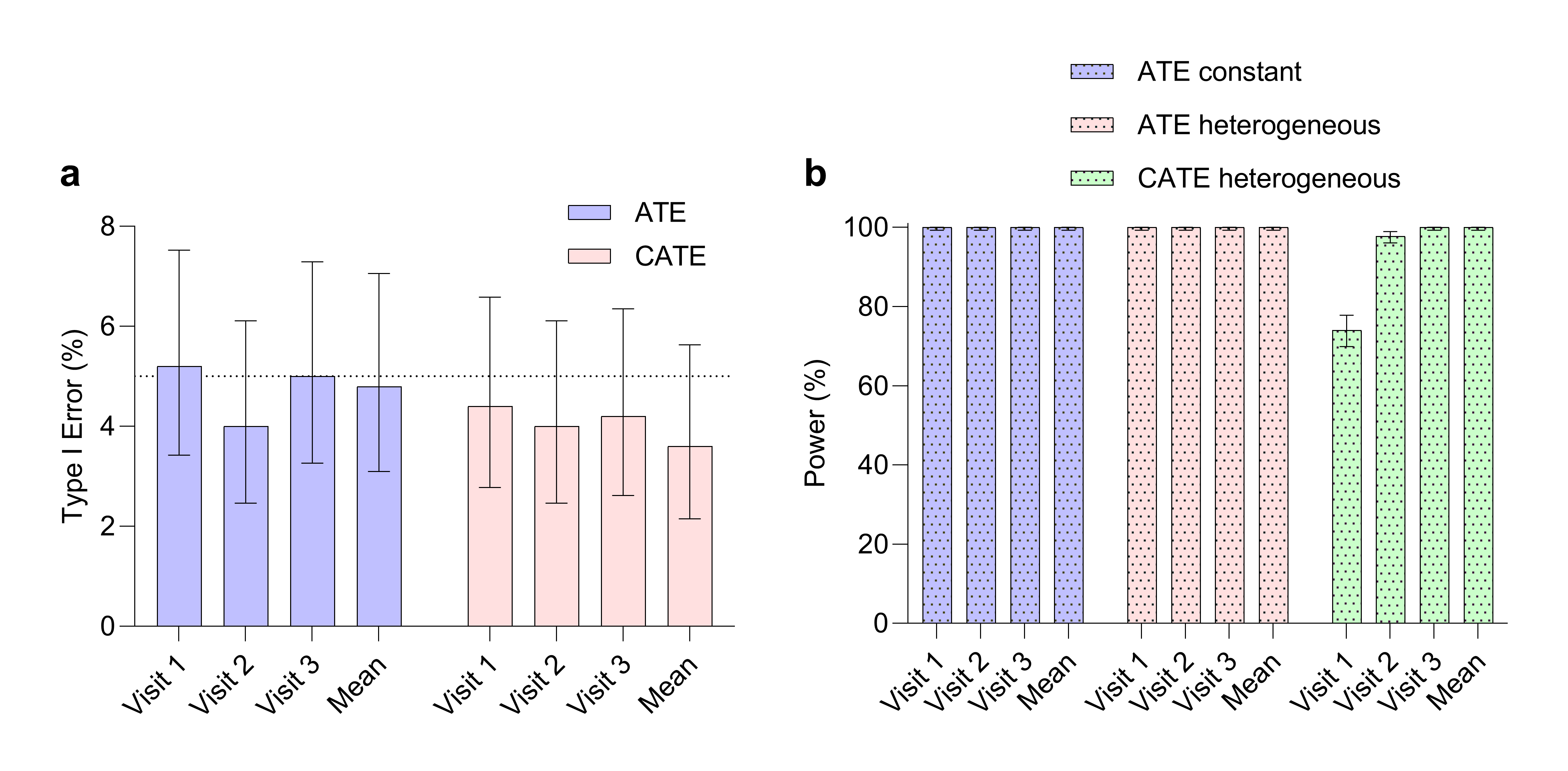}
    \caption{\textbf{Type I error and power.}}
    \label{fig:sim:typeIerror_power}
\end{figure}

\begin{figure}[H]
    \centering
    \includegraphics[width=0.95\linewidth]{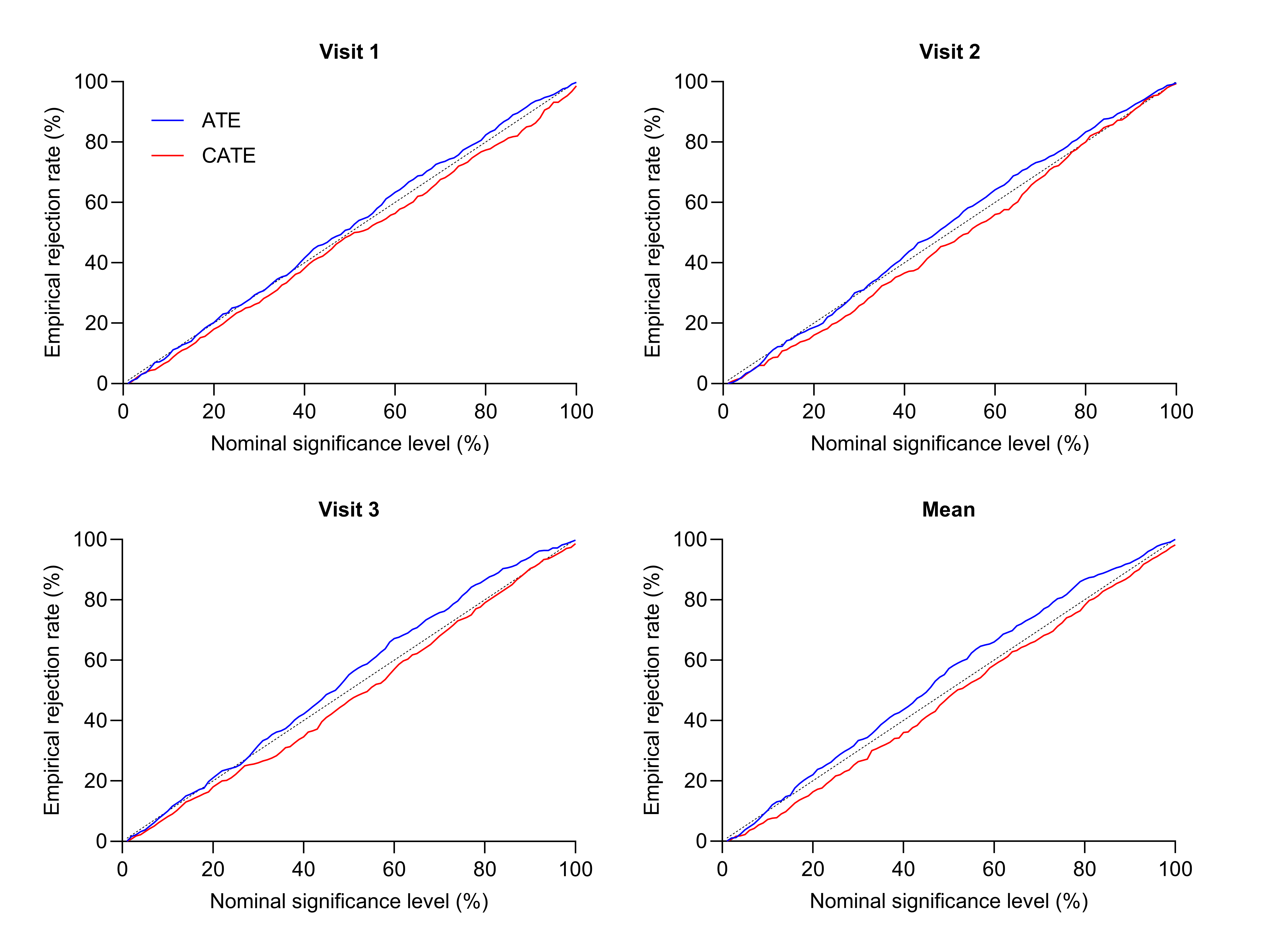}
    \caption{\textbf{Calibration plots under the null hypotheses.} Curves below the diagonal indicate conservative tests. The CRTs are calibrated at nominal significance levels below 10\%.}
    \label{fig:sim:calibration}
\end{figure}

\newpage
\subsection{Extra Semi-Synthetic Data Results}

\begin{figure}[H]
    \centering
    \includegraphics[width=0.9\linewidth]{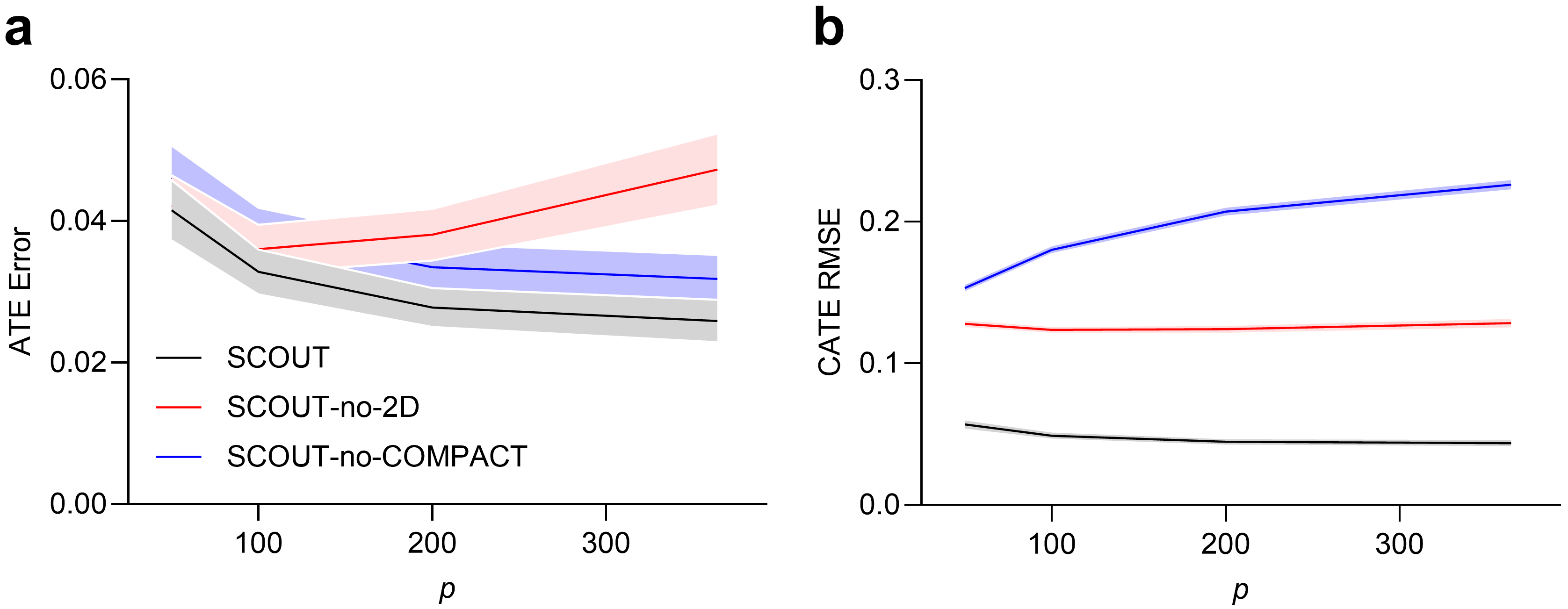}
    \caption{\textbf{Ablation results using CATIE-based semi-synthetic data.} SCOUT consistently achieved lower (a) ATE error and (b) CATE RMSE than its ablated variants, with the advantage in ATE estimation becoming more pronounced as the number of baseline proxies increased.}
    \label{fig:CATIE:ablations}
\end{figure}

\begin{figure}[H]
    \centering
    \includegraphics[width=0.45\linewidth]{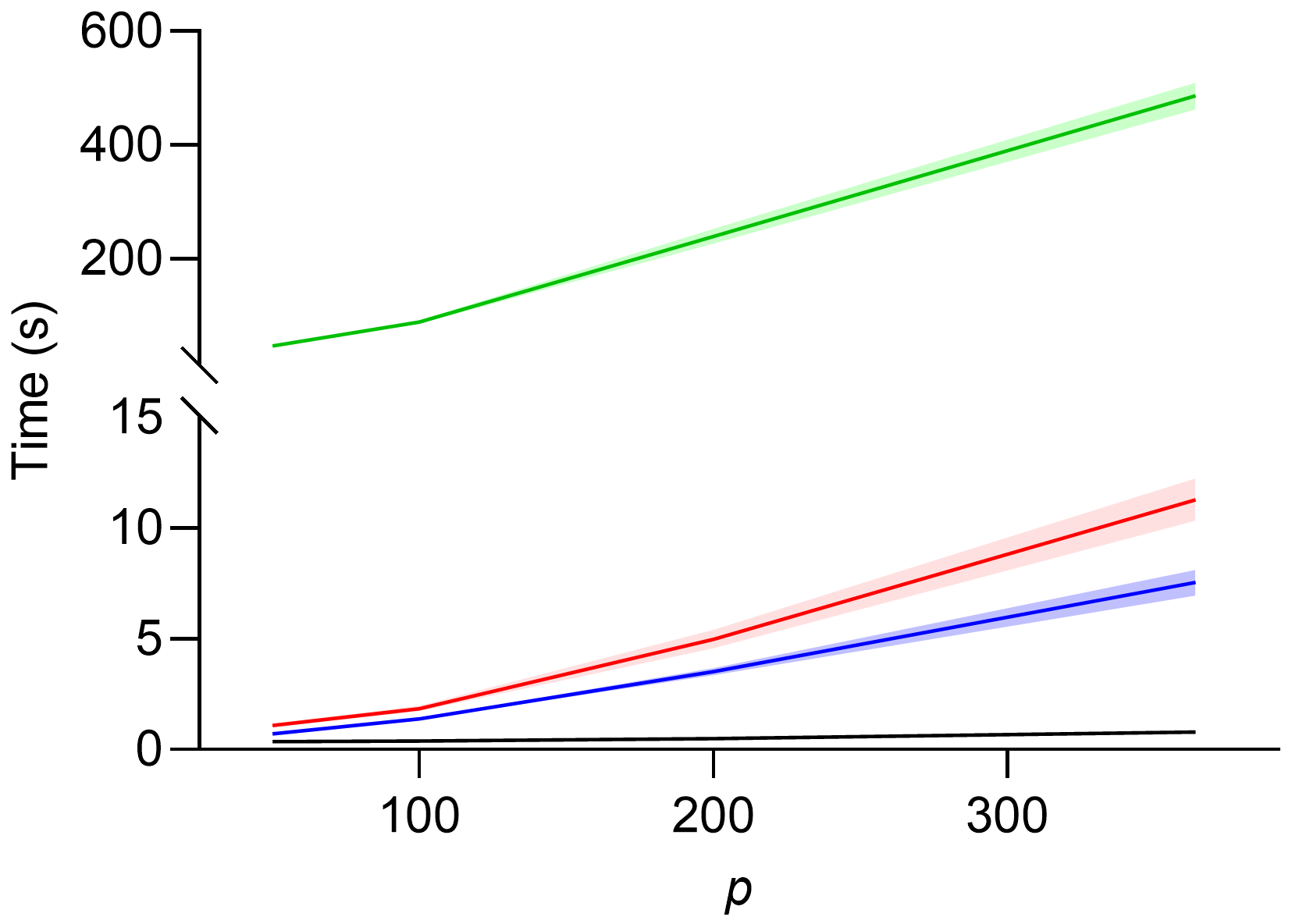}
    \caption{\textbf{Computational runtime in CATIE-based semi-synthetic experiments.}}
    \label{fig:CATIE:time}
\end{figure}

\begin{figure}[H]
    \centering
    \includegraphics[width=0.9\linewidth]{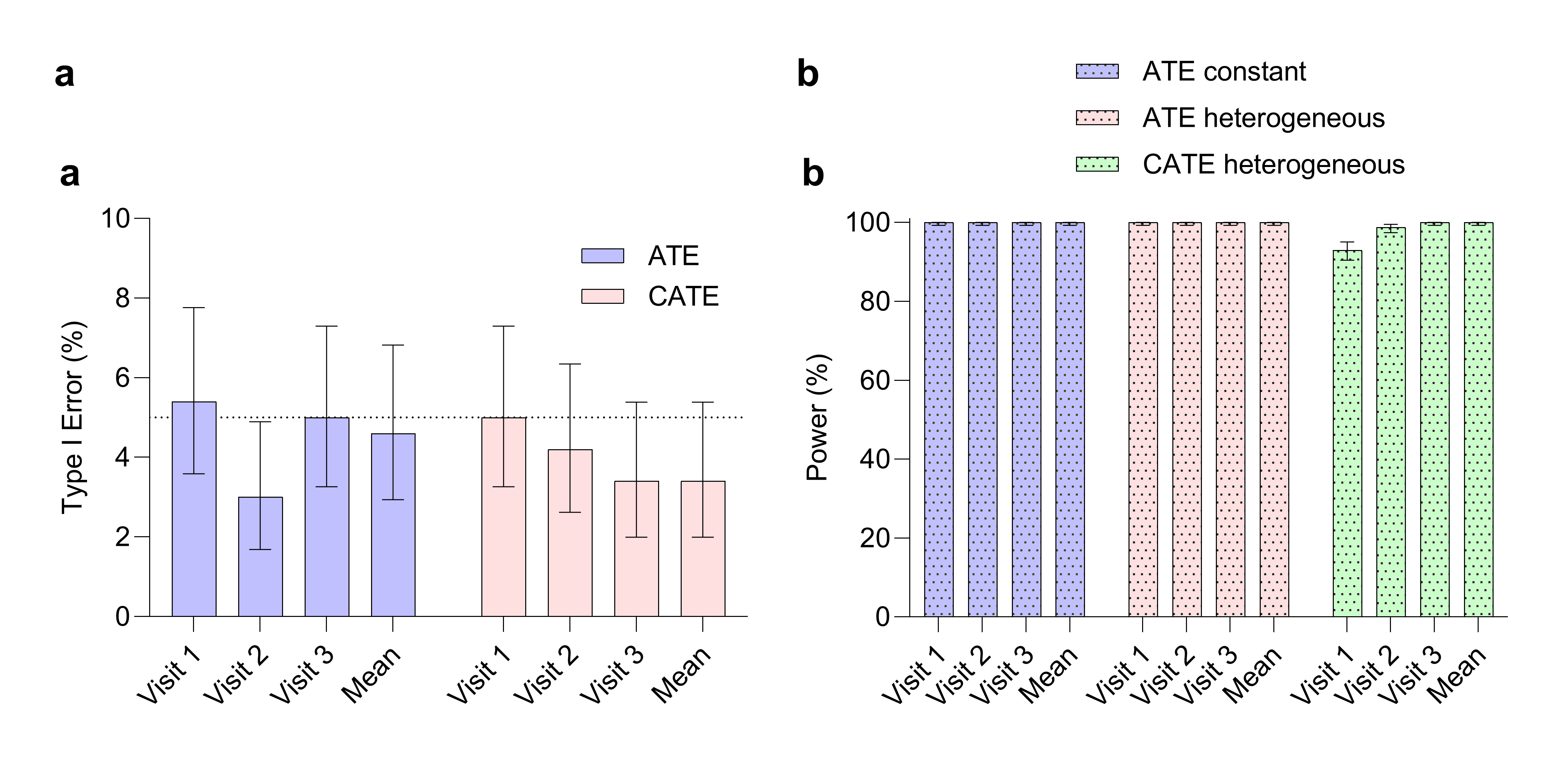}
    \caption{\textbf{Type I error and power in CATIE-based semi-synthetic experiments.}}
    \label{fig:sim:catie_typeIerror_power}
\end{figure}

\begin{figure}[H]
    \centering
    \includegraphics[width=0.95\linewidth]{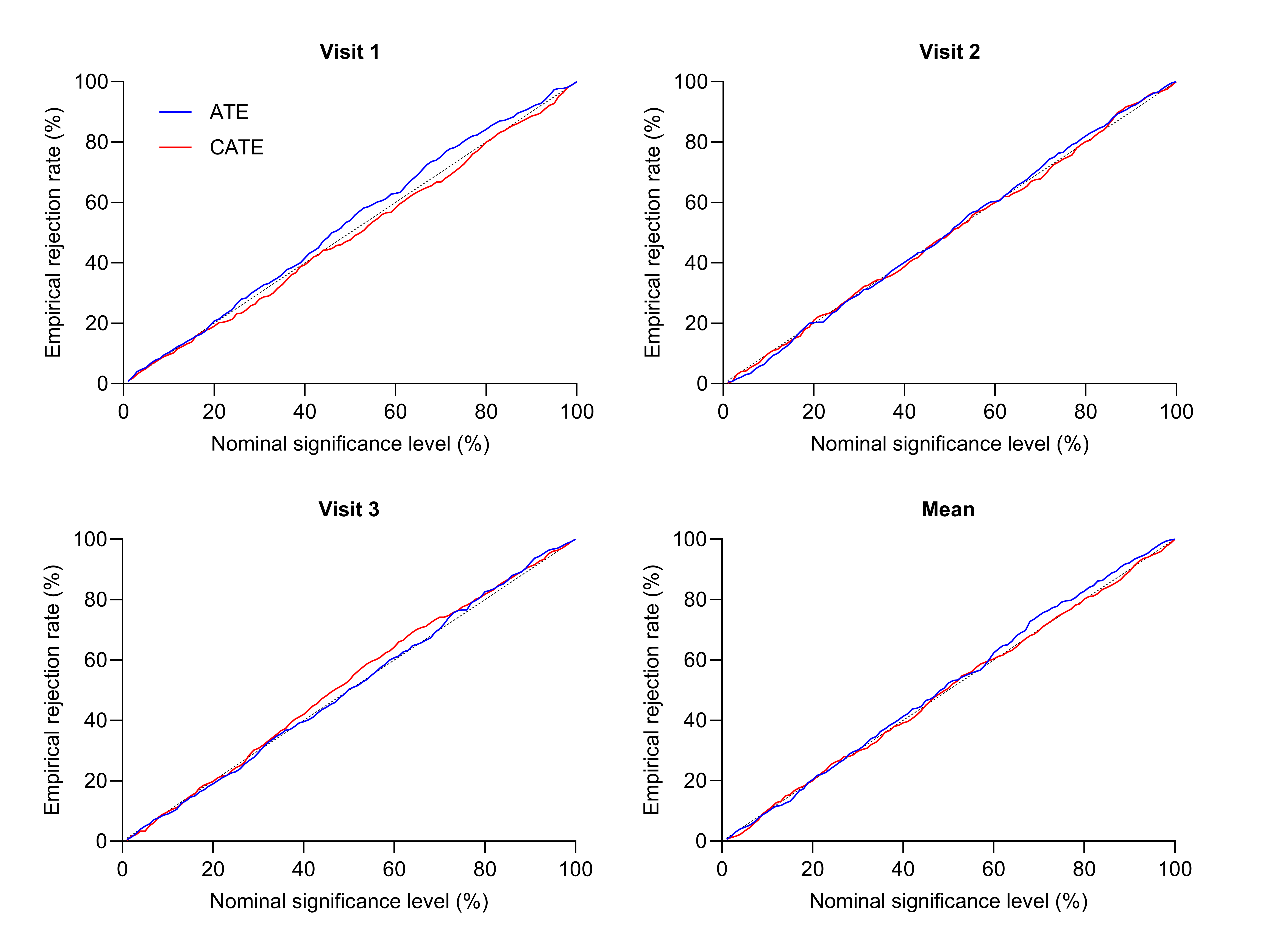}
    \caption{\textbf{Calibration plots in CATIE-based semi-synthetic experiments under the null hypotheses.}}
    \label{fig:sim:catie_calibration}
\end{figure}

\newpage \subsection{Extra Real Application Results}

\begin{figure}[H]
    \centering
    \includegraphics[width=0.8\linewidth]{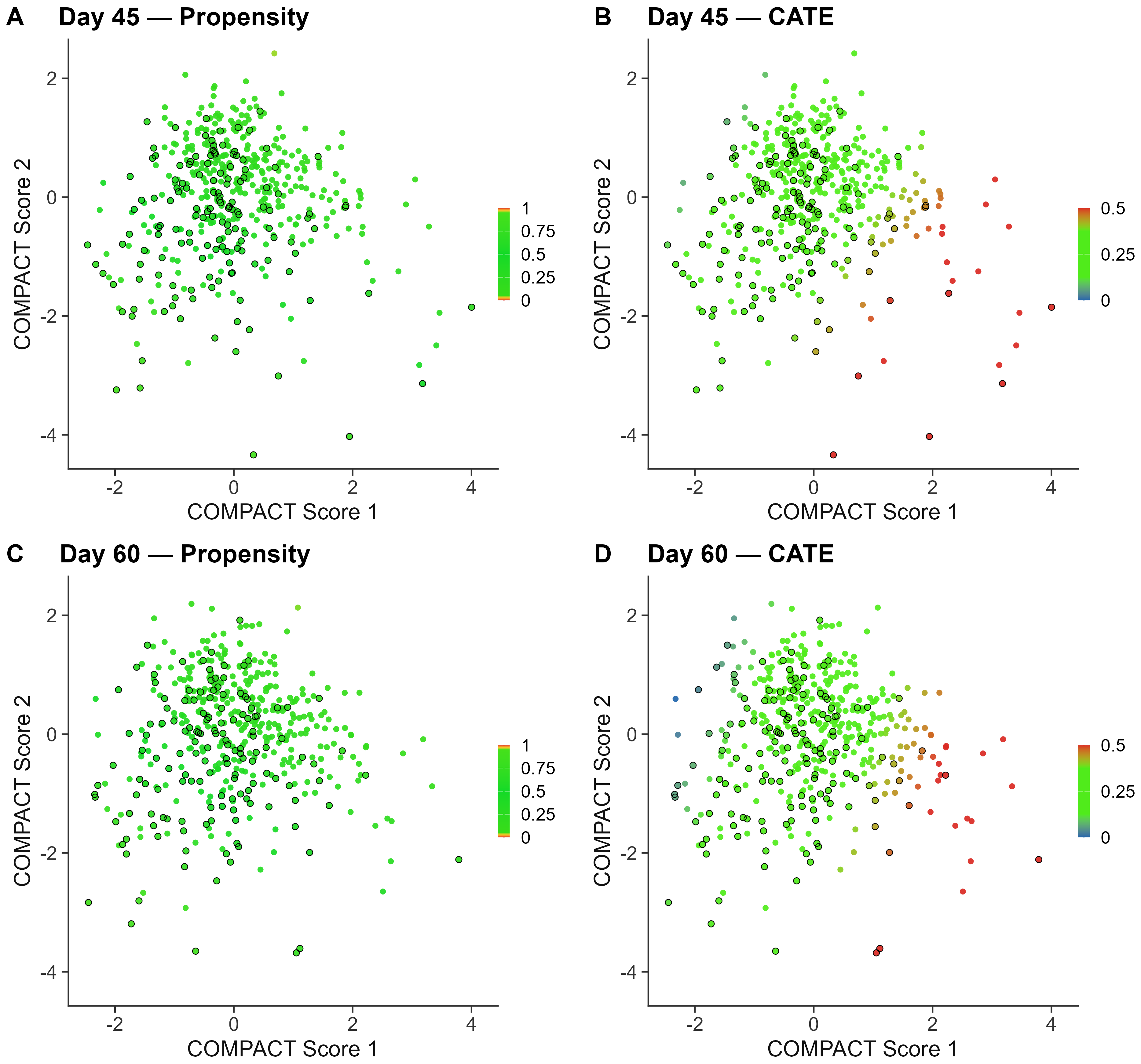}
    \caption{\textbf{Visualization of SCOUT propensity scores and estimated CATEs over the two-dimensional adjustment space on days 45 and 60.}
    (a,c) Estimated propensity scores conditional on the two-dimensional adjustment scores at each time point, showing substantial treatment overlap without extreme propensity scores.
    (b,d) Estimated CATEs over the corresponding adjustment spaces. Green indicates CATEs within one standard deviation of the mean, while blue and red indicate progressively lower and higher CATEs, respectively.
    The plots reveal distinct regions of elevated treatment effects (red), consistent with the treatment-effect heterogeneity detected by the CRT.
    Each point represents a patient, with black outlines identifying lithium-treated patients ($T=1$).}
    \label{fig:res:STEPBD}
\end{figure}

\begin{figure}[H]
    \centering
    \includegraphics[width=0.8\linewidth]{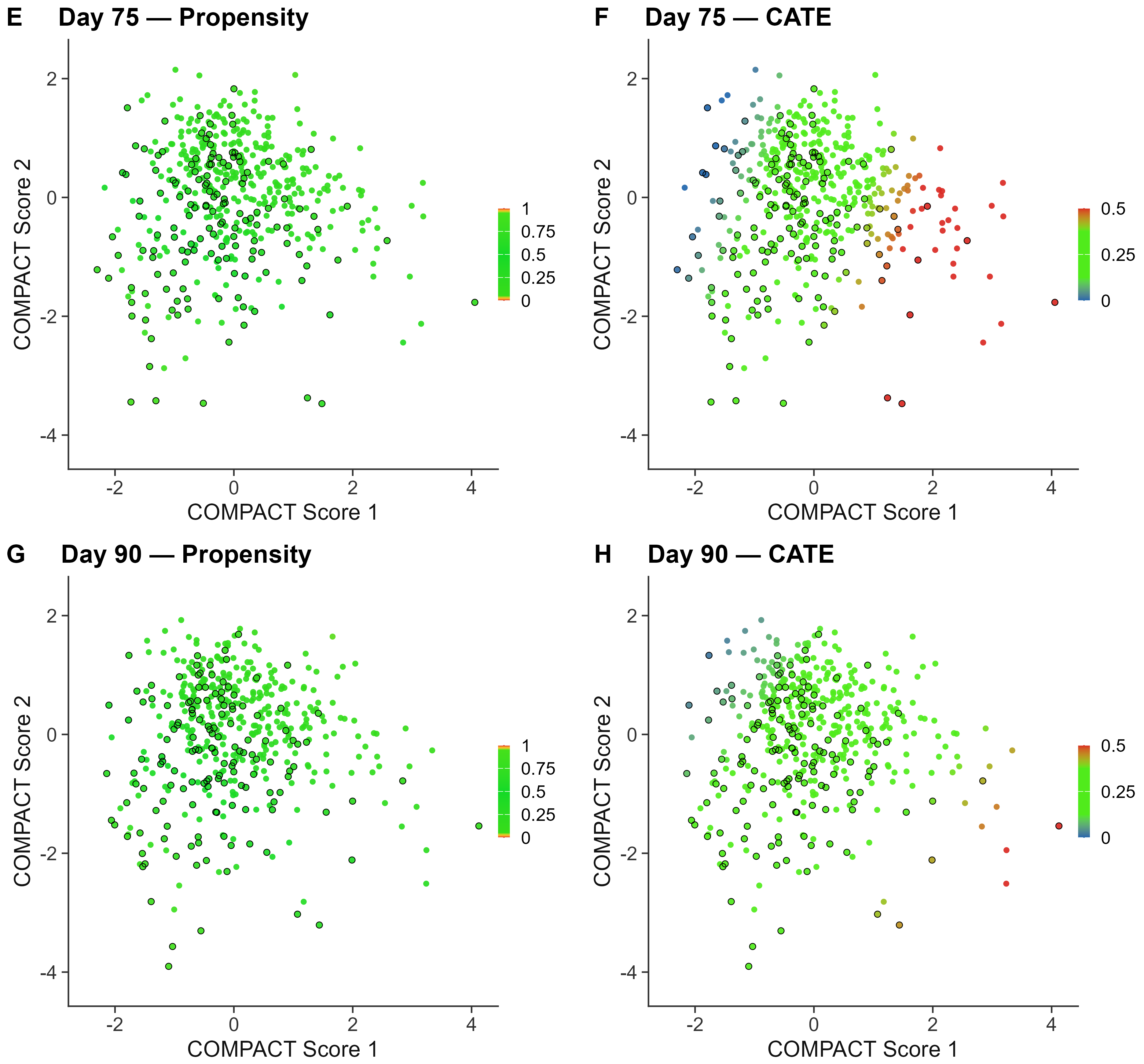}
    \caption{\textbf{Visualization of SCOUT propensity scores and estimated CATEs over the two-dimensional adjustment space on days 75 and 90.}}
    \label{fig:res:STEPBD2}
\end{figure}

\end{document}